\documentclass[journal]{IEEEtran}
\usepackage{bm}
\usepackage{amsfonts}
\usepackage{amsmath}

\newtheorem{definition}{Definition}

\usepackage{graphicx}  
\usepackage{epstopdf}
\usepackage{multirow}
\usepackage{booktabs}

\begin{document}

\title{Spatio-Temporal Correlation Analysis of Online Monitoring Data for Anomaly Detection and Location in Distribution Networks}

\author{ Xin Shi$^1$,~\IEEEmembership{Student Member,~IEEE}, Robert Qiu$^{1,2}$,~\IEEEmembership{Fellow,~IEEE}, \\
Zenan Ling$^1$, Fan Yang$^1$, Haosen Yang$^1$, Xing He$^1$
\thanks{This work was partly supported by National Key R \& D Program of No. 2018YFF0214705, NSF of China No. 61571296 and (US) NSF Grant No. CNS-1619250.

$^1$ Department of Electrical Engineering, Center for Big Data and Artificial Intelligence, State Energy Smart Grid Research and Development Center, Shanghai Jiaotong University, Shanghai 200240, China. (e-mail: dugushixin@sjtu.edu.cn; rcqiu@sjtu.edu.cn; ling\_zenan@163.com; 1164431011@qq.com; 31910019@sjtu.edu.cn; hexing\_hx@126.com)

$^2$ Department of Electrical and Computer Engineering,Tennessee Technological University, Cookeville, TN 38505, USA. (e-mail:rqiu@tntech.edu)

}
}

\maketitle

\begin{abstract}
 The online monitoring data in distribution networks contain rich information on the running states of the networks. By leveraging the data, this paper proposes a spatio-temporal correlation analysis approach for anomaly detection and location in distribution networks. First, spatio-temporal matrix for each feeder line in a distribution network is formulated and the spectrum of its covariance matrix is analyzed. The spectrum is complex and exhibits two aspects: 1) bulk, which arises from random noise or fluctuations and 2) spikes, which represents factors caused by anomaly signals or fault disturbances. Then, by connecting the estimation of the number of factors to the limiting empirical spectral density of covariance matrices of residuals, the spatio-temporal parameters are accurately estimated, during which free random variable techniques are used. Based on the estimators, anomaly indicators are designed to detect and locate the anomalies by exploring the variations of spatio-temporal correlations in the data. The proposed approach is sensitive to the anomalies and robust to random fluctuations, which makes it possible for detecting early anomalies and reducing false alarming rate. Case studies on both synthetic data and real-world online monitoring data verify the effectiveness and advantages of the proposed approach.
\end{abstract}


\IEEEpeerreviewmaketitle

\section{Introduction}
\label{section: Introduction}

\IEEEPARstart{T}{his} paper is driven by the need of anomaly detection and location using online monitoring data in a distribution network. The anomalies caused by some fault disturbances may present intermittent, asymmetric, and sporadic spikes, which are random in magnitude and could involve sporadic bursts as well, and exhibit complex, nonlinear, and dynamic characteristics \cite{jaafari2007underground}. What's more, with numerous branch lines and changeable network topology, it is questionable that traditional model-based approaches are capable of fully and accurately detecting and locating the anomalies in the distribution network, because they are usually based on certain assumptions and simplifications.

With significant deployment of online monitoring devices in distribution networks, a large amount of data is collected. In order to leverage the data, many advanced analytics have been developed in recent years. For example, in \cite{ma2003time}, one-class support vector machines (SVMs) is proposed for time-series novelty detection. In \cite{Dasgupta2013Real}, time-series voltage data from online monitoring system is used to compute Lyapunov component to estimate voltage stability. In \cite{xie2014dimensionality}, dimensionality of synchrophasor data is analyzed and a PCA-based dimension reduction algorithm is developed for early event detection. In \cite{malhotra2015long}, stacked long short term memory (LSTM) networks are developed for time-series anomaly detection. In \cite{Chu2016Massive}, by modeling streaming PMU data as random matrix flow, an algorithm based on multiple high dimensional covariance matrix tests is developed for system state estimation. In \cite{liu2018anomaly}, structured neural networks are proposed for anomaly detection in manufacturing systems.

For a system with multiple measurement devices installed, the multi-dimensional data collected through them contains rich information on the system states. In terms of data structure, spatio- (cross-) and temporal (auto) correlation should be considered when analyzing the system states. Then several open questions are raised, for example:
1) What is the spatio-temporal correlation of the data?
2) How to characterize or measure the spatio-temporal correlation of the data?
3) What is the relationship between the spatio-temporal correlation of the data and the state of the system?
It is questionable for the conventional model-based methods to model the complex system, let alone addressing the above questions.

Factor models are important tools for reducing the dimensionality and extracting the relevant information in analyzing high-dimensional data, which have been well studied in statistics and econometrics. In \cite{kapetanios2010testing}, factor models are used for modeling a large number of economic variables, and the structure of residuals is exploited for estimating the number of factors. In \cite{Harding2013Estimating}, restrictions on the structure of residuals are imposed to improve the performance of estimating weak factors in asset pricing. In \cite{Ahn2013Eigenvalue}, a new estimator is proposed for determining the number of factors by maximizing the radio of two adjacent eigenvalues, which has good finite sample properties on Monte Carlo simulation data. In \cite{pelger2019large}, factor models are successfully applied to financial high-frequency data analysis. In \cite{Yeo2016Random}, a new approach to estimate high-dimensional factor models is proposed. The proposed approach can effectively capture the structural information of the data and outperforms other known methods. Considering the structure of the real-world online monitoring data is complex and cannot be trivially dissected by simple techniques, it is meaningful to apply factor models for the real data analysis in distribution networks.

In this paper, based on exploring the spatio-temporal correlation of the data amongst multiple monitoring devices in a distribution network, a new approach for anomaly detection and location is proposed. It leverages the spatio-temporal similarities amongst the data, and realizes anomaly detection and location by measuring the variations of the spatio-temporal correlation of the data. The main advantages of the proposed approach can be summarized as follows:
1) It is a purely data-driven approach without requiring too much prior knowledge on the complex topology of the network.
2) It is sensitive to the variation of the spatio-temporal correlation of the online monitoring data, which makes it possible for detecting the anomalies in an early phase. Because the correlation of the data usually changes immediately once an anomaly occurs.
3) It is theoretically and experimentally justified that the proposed approach is robust to random fluctuations and measuring errors in the data, which can help reduce the false alarming rate.
4) The approach is suitable for both online and offline analysis.

The rest of this paper is organized as follows. Section \ref{section: problem} analyzes the empirical spectrum distribution of the online monitoring data and the anomaly detection problem is formulated as the estimation of spatio-temporal parameters. In Section \ref{section: anomaly_detection}, the anomaly detection and location approach based on spatio-temporal correlation analysis is proposed and discussed. Both synthetic data from IEEE 33-bus, 57-bus test system and real-world online monitoring data from a grid are used to validate the effectiveness and advantages of the proposed approach in Section \ref{section: case_study}. Conclusions are presented in Section \ref{section: conclusion}.

\section{Problem Formulation}
\label{section: problem}
In this section, the empirical spectrum distribution (ESD) of the covariance matrix of the online monitoring data in a distribution network under both normal and abnormal feeder operating states is first analyzed. Then, the residuals obtained by subtracting principal components from the real data are formulated and discussed. The anomaly detection and location problem is connected to the estimation of spatio-temporal parameters.

\subsection{Empirical Spectrum Distribution of the Online Monitoring Data}
\label{section: esd}
\begin{figure}[!t]
\centering
\begin{minipage}{4.1cm}
\centerline{
\includegraphics[width=1.9in]{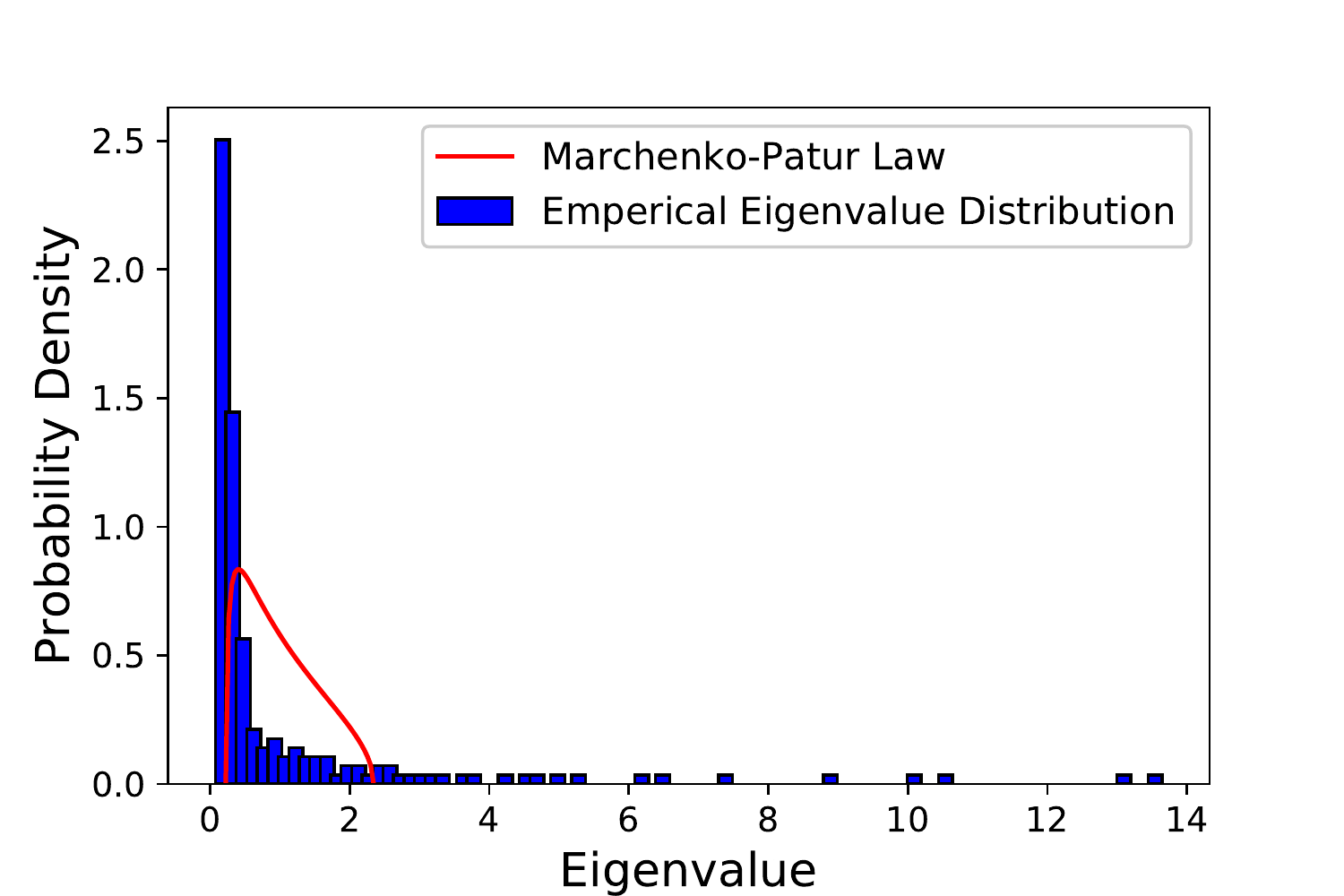}
}
\parbox{5cm}{\small \hspace{1.2cm}(a) Normal state }
\end{minipage}
\hspace{0.2cm}
\begin{minipage}{4.1cm}
\centerline{
\includegraphics[width=1.9in]{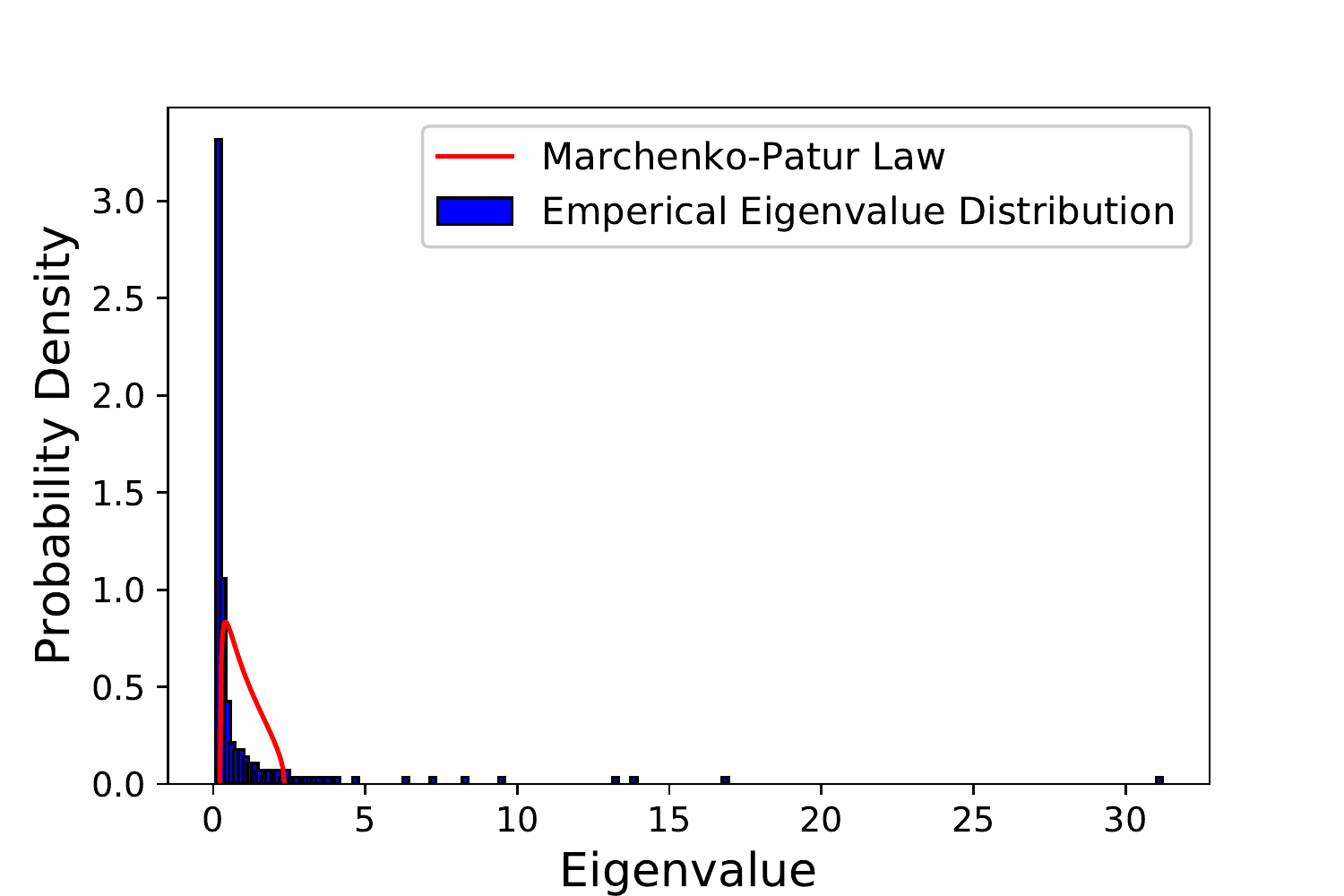}
}
\parbox{5cm}{\small \hspace{1.2cm}(b) Abnormal state }
\end{minipage}
\caption{The ESD of the covariance matrix of $\bf\hat R$ and its comparison with the theoretical M-P law under both normal and abnormal feeder operating states.}
\label{fig:data_example}
\end{figure}

We apply the Marchenko-Pastur law (M-P law) \cite{marvcenko1967distribution} for the online monitoring data from a distribution network. Definition about the M-P law can be found in Appendix \ref{section: mp_law}. Figure \ref{fig:data_example_org} shows three-phase voltage magnitude curves collected from one feeder line. The feeder contained $63$ distribution transformers in total. On the low voltage side of each transformer, one online monitoring device was installed, through which three-phase voltage measurement can be obtained. The data were sampled every 15 minutes and the sampling time was from 2017/3/1 00:00:00 to 2017/3/14 23:45:00, thus a $189\times 1344$ data set was formulated. Let $\bf R$ be a $189\times 672$ moving window on the data set and we convert $\bf R$ into the standard form $\bf\hat R$ through
\begin{equation}
\label{Eq:standardize}
\begin{aligned}
  {\hat r_{ij}} = \left( {{r_{ij}} - \mu \left( {{{\bf r}_i}} \right)} \right) \times \frac{{\sigma \left( {{{\hat {\bf r}}_i}} \right)}}{{\sigma \left( {{{\bf r}_i}} \right)}} + \mu \left( {{{\hat {\bf r}}_i}} \right)
\end{aligned},
\end{equation}
where ${\bf r}_i=(r_{i1},r_{i2},...)$, $\mu ({\bf\hat r}_i)=0$, and $\sigma ({\bf\hat r}_i)=1$. The covariance matrices of $\bf\hat R$ corresponding to the normal and abnormal data windows in Figure \ref{fig:data_example_org} are calculated and the ESDs with top $5$ factors removed are shown in Figure \ref{fig:data_example}.
\begin{figure}[!t]
\centerline{
\includegraphics[width=3.0in]{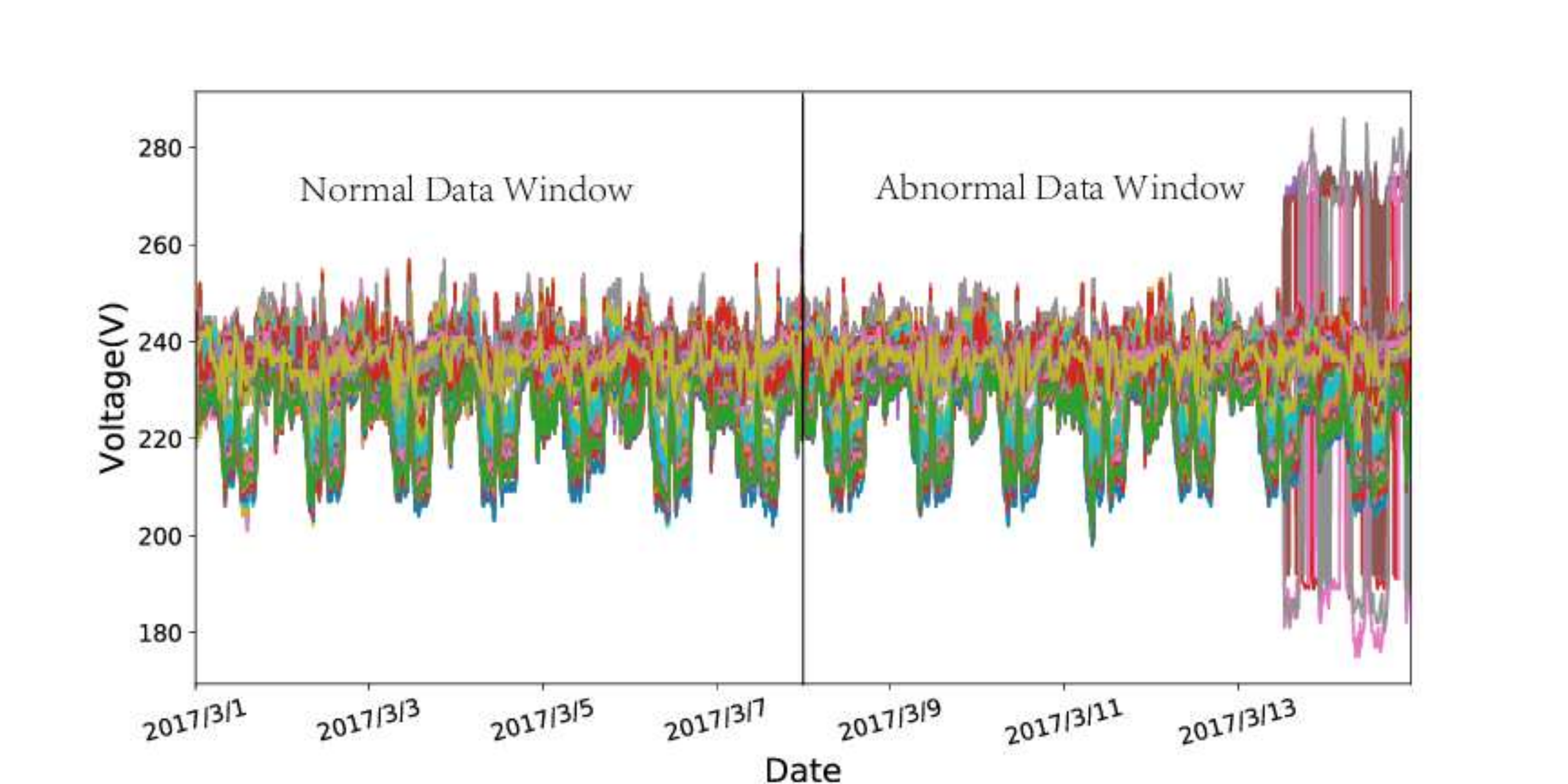}
}
\caption{Three-phase voltage magnitude curves.}
\label{fig:data_example_org}
\end{figure}

From Figure \ref{fig:data_example}, It can be observed that the spectrum of covariance matrix of $\bf\hat R$ typically exhibits two aspects: bulk (i.e., the blue bars) and spikes (i.e., the deviating eigenvalues). The bulk arises from random noise or fluctuations and the spikes are mainly caused by fault disturbances. It is noted that the spectrum can not be fit by the M-P law whether the feeder line operates in normal or abnormal state, but the region of the bulk and the size of the spikes are different when the feeder line operates in different states. Therefore, the spectrum can not be trivially dissected by using the M-P law, and we must consider a new approach to depict the complex spectrum for detecting anomalies more accurately.
\subsection{Residual Formulation and Discussion}
\label{section: residual}
\begin{figure}[htb]
\centering
\begin{minipage}{4.1cm}
\centerline{
\includegraphics[width=1.9in]{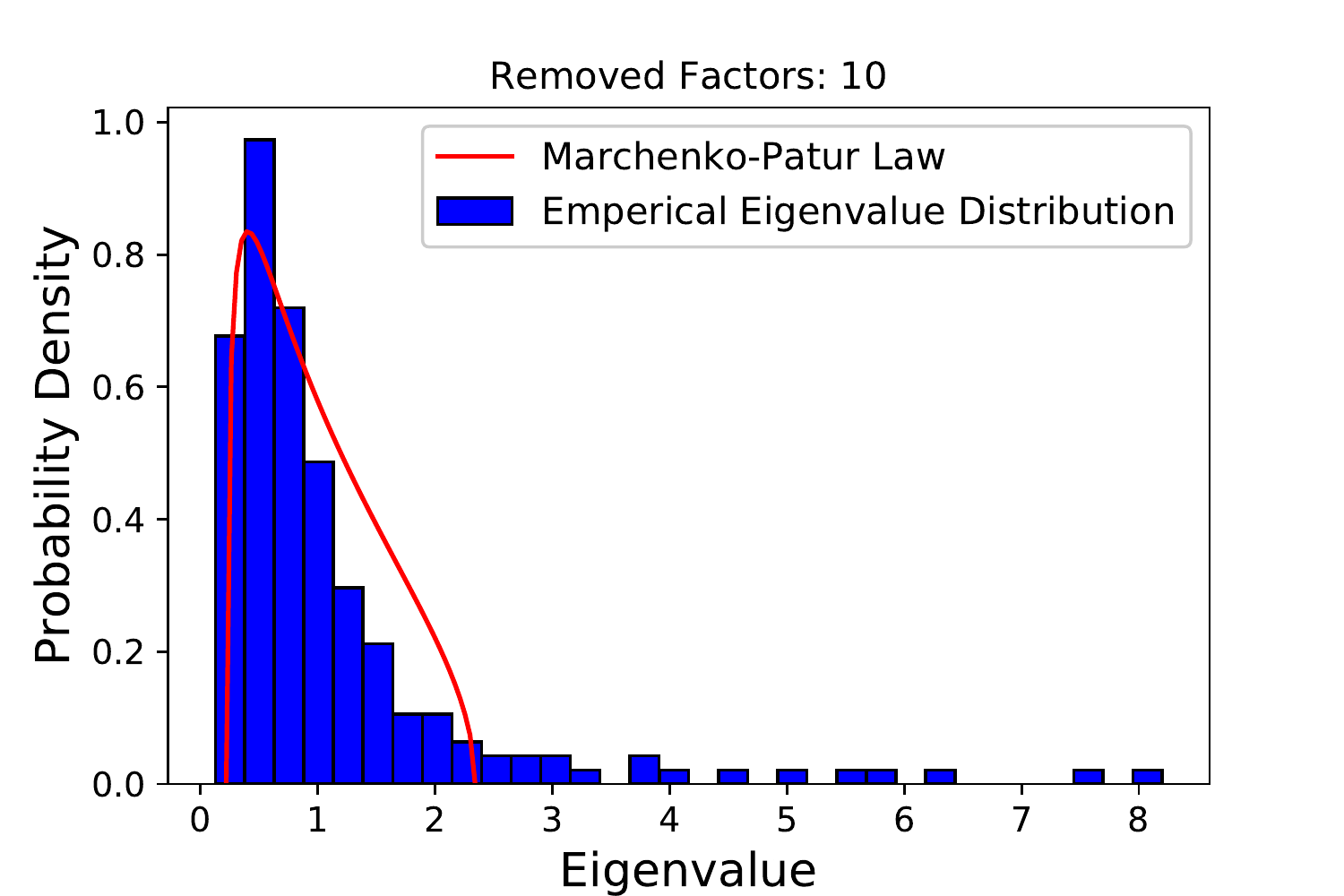}
}
\parbox{5cm}{\small \hspace{1.8cm}(a) }
\end{minipage}
\hspace{0.2cm}
\begin{minipage}{4.1cm}
\centerline{
\includegraphics[width=1.9in]{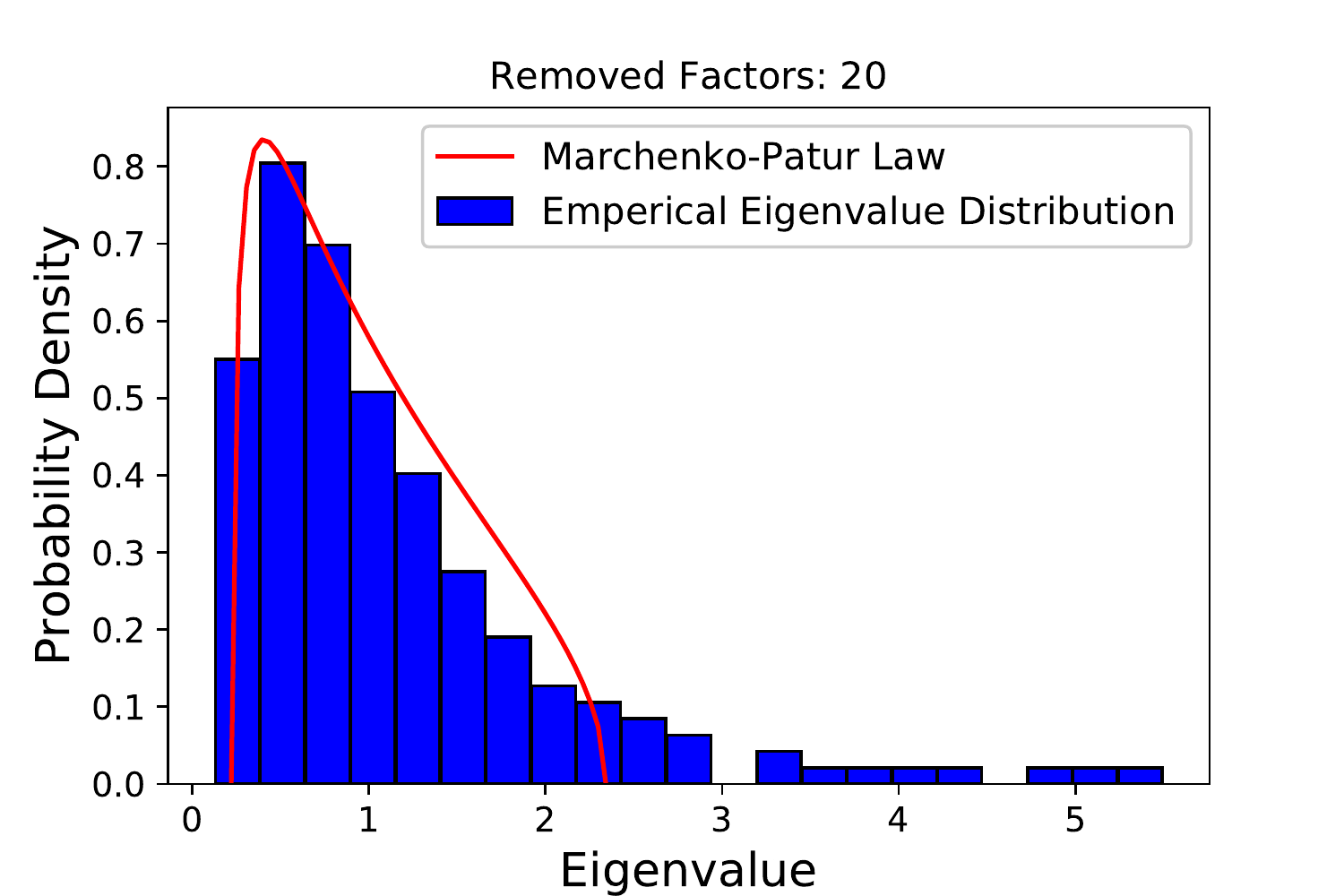}
}
\parbox{5cm}{\small \hspace{1.8cm}(b) }
\end{minipage}
\hspace{0.2cm}
\begin{minipage}{4.1cm}
\centerline{
\includegraphics[width=1.9in]{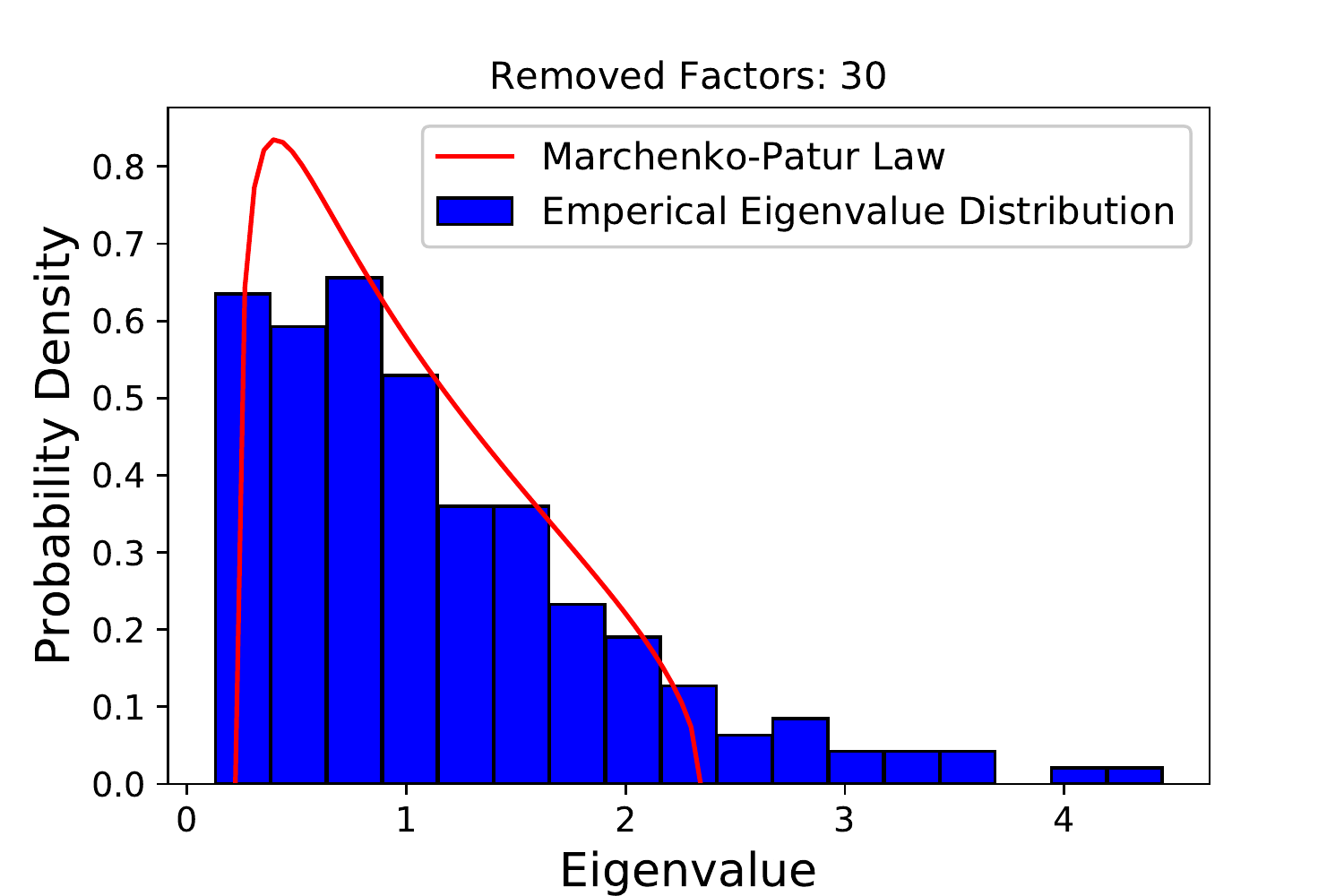}
}
\parbox{5cm}{\small \hspace{1.8cm}(c) }
\end{minipage}
\hspace{0.2cm}
\begin{minipage}{4.1cm}
\centerline{
\includegraphics[width=1.9in]{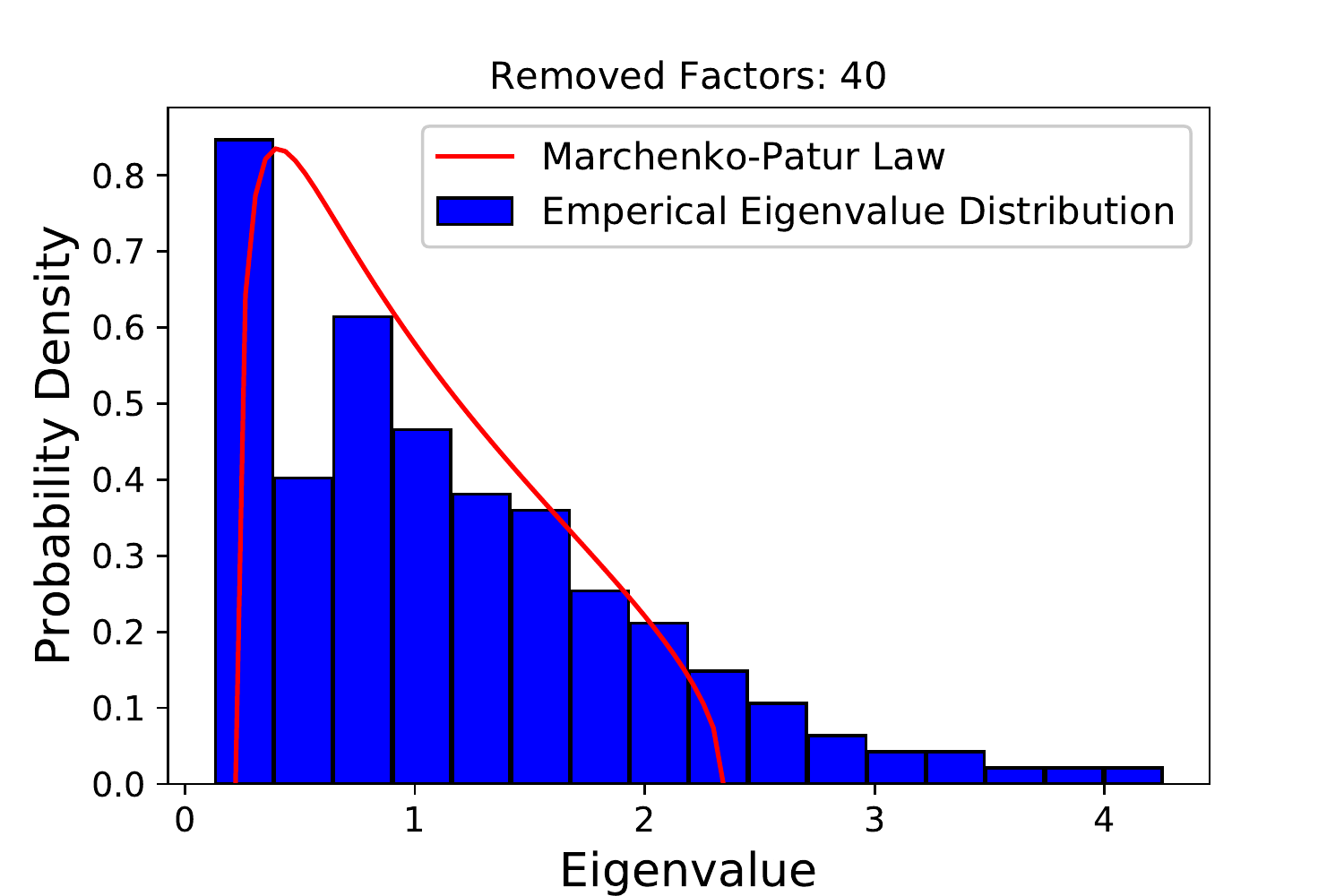}
}
\parbox{5cm}{\small \hspace{1.8cm}(d) }
\end{minipage}
\caption{No matter how many factors are removed, the ESD of the covariance matrix of the residuals from the real-world online monitoring data does not converge to the M-P law.}
\label{fig:factor_remove}
\end{figure}

From subsection \ref{section: esd}, the spectrum of the covariance matrix of $\bf\hat R$ inspires us to decompose the real-world online monitoring data into systematic components (factors) and idiosyncratic noise (residuals). Assume matrix $\bf R$ is of $N$ measurements and $T$ observations, thus a factor model regarding $\bf R$ can be written as
\begin{equation}
\label{Eq:factor_model}
\begin{aligned}
  {\bf R}= LF + U
\end{aligned},
\end{equation}
where $L$ is an $N\times p$ matrix of factor loadings, $F$ is a $p\times T$ matrix of factors, $p$ is the number of factors, and $U$ is an $N\times T$ matrix of residuals. For the real-world online monitoring data, the ESD of the covariance matrix of the residuals does not fit to the M-P law, no matter how many factors are removed, as is shown in Figure \ref{fig:factor_remove}.

In order to estimate the spectrum of the real residuals, we connect the estimation of the number of factors to the limiting ESD of the covariance matrix of $U$. Assume there are cross- and auto-correlated structures in $U$, then it can be denoted as $U = {A_{N}^{1/2}}S{B_{T}^{1/2}}$. The covariance matrix of $U$ is written as $\Sigma = \frac{1}{T}UU^{T} = \frac{1}{T}{A_{N}^{1/2}}S{B_T}{S^T}{A_{N}^{1/2}}$, where $S$ is an $N\times T$ matrix, $A_N$ and $B_T$ are $N\times N$ and $T\times T$ symmetric non-negative definite matrices, respectively representing cross- and auto- covariances. The structures of $A_N$ and $B_T$ are restricted so that they can be determined by the parameter set $\theta$ (i.e., $\theta = (\theta_{A_N}, \theta_{B_T})$). For example, a simple case is that each residual has the same cross-correlation with parameter $\beta$ and an exponentially decaying auto-correlation with parameter $\tau$, i.e., $A_N=\{({A_N})_{ii}=1, ({A_N})_{ij,i\neq j}=\beta, i,j=1,\cdots,N\}$, $B_T=\{({B_T})_{st}=exp(-|s-t|/\tau), s,t=1,\cdots,T \}$. The objective of our estimation method is to match the eigenvalue distribution of $\Sigma$ to that of the covariance matrix of residuals constructed from real data. The latter is controlled by the number of removing factors (i.e., parameter $p$), and the former is determined by the parameter set $\theta$. We can search $p$ and $\theta$, such that the spectral distance between the model (i.e., $\rho_{model}(\theta)$) and real data (i.e., $\rho_{real}(p)$) is minimized. A difficulty in the implementation is the calculation of $\rho_{model}(\theta)$ for general $A_N$ and $B_T$. Therefore, we make two assumptions here for simplifying the modeling for $A_N$ and $B_T$.

$Assumption\; 1$: The cross-correlations of the real residual $U^{(p)}$ are effectively eliminated by removing $p$ factors, thus, $A_N \approx I_{N\times N}$.

$Assumption\; 2$: The auto-correlations of the real residual $U^{(p)}$ are exponentially decreasing, thus, $\{ B_T \}_{ij} = b^{|i-j|}$, with $|b|<1$.

From $Assumption\; 1$ and $2$, the calculation of $\rho_{model}(\theta)$ is replaced by $\rho_{model}(b)$, and the minimization problem has only two parameters, i.e., $p$ and $b$. The two parameters effectively characterize the features of the spectrum from real data: $p$ controls the range of spikes, and $b$ reflects the shape of the bulk.
Combining the spectrum analysis results in Section \ref{section: esd}, $p$ and $b$ can be used as the basis for detecting the anomalies in distribution networks.
\section{Anomaly Detection and Location in Distribution Networks}
\label{section: anomaly_detection}
Based on the discussions above, by using the online monitoring data in distribution networks, a spatio-temporal correlation analysis approach is proposed for anomaly detection and location. In this section, the estimation method of factor models in equation (\ref{Eq:factor_model}) is illustrated in detail, in which FRV techniques \cite{burda2010random} are used to calculate the modeled spectral density. Then, specific steps of the proposed anomaly detection and location approach are given, and the advantages of the approach are systematically analyzed. Finally, more discussions about the proposed approach are presented.
\subsection{Factor Model Estimation}
\label{section: factor_model_estimation}
From Section \ref{section: residual}, we can estimate $p$ and $b$ by minimizing the spectral distance between the model and the real data, which is stated as
\begin{equation}
\label{Eq:distance}
\begin{aligned}
  \{{\hat p},{\hat b}\} = \text{arg}\min\limits_{p,b} \mathcal{D}(\rho_{real}(p), \rho_{model}(b))
\end{aligned},
\end{equation}
where $\rho_{real}(p)$ represents the ESD of the covariance matrix of the residuals constructed by removing $p$ factors from the real data, $\rho_{model}(b)$ is the limiting spectral density of the modeled covariance matrix characterized by parameter $b$, and $\mathcal{D}$ is the spectral distance measure.

In order to obtain $\rho_{real}(p)$, we firstly obtain the residuals by removing $p$ largest principal components from the real data. Because for high dimensional data, principal components can approximately mimic all true factors \cite{Stock2002Forecasting}. Considering the factor model in equation (\ref{Eq:factor_model}), the $p-$level residual ${\hat U}^{(p)}$ is calculated by
\begin{equation}
\label{Eq:residual}
\begin{aligned}
  {\hat U}^{(p)} = R - {\hat L}^{(p)}{\hat F}^{(p)}
\end{aligned},
\end{equation}
where ${\hat F}^{(p)}$ is an $p\times T$ matrix of $p$ principal components from correlation matrix of $R$, ${\hat L}^{(p)}$ is an $N\times p$ matrix of factor loadings, estimated by multivariate least squares regression of R on ${\hat F}^{(p)}$, namely
\begin{equation}
\label{Eq:factor_loading}
\begin{aligned}
  {\hat L}^{(p)} = R*inv({{\hat F}^{(p)}})
\end{aligned},
\end{equation}
where $inv()$ denotes the pseudo-inverse operation. The covariance matrix of ${\hat U}^{(p)}$ is calculated as
\begin{equation}
\label{Eq:covariance_real}
\begin{aligned}
  {\Sigma}_{real}^{(p)} = \frac{1}{T}{{\hat U}^{(p)}}{\hat U}^{{(p)}^{T}}
\end{aligned},
\end{equation}
and $\rho_{real}(p)$ is the ESD of ${\Sigma}_{real}^{(p)}$.

Then we calculate $\rho_{model}(b)$ by using FRV techniques. For the autoregressive model $U$:
\begin{equation}
\label{Eq:autoregressive_model}
\begin{aligned}
  U_{it}=bU_{i,t-1}+\varepsilon_{it}
\end{aligned},
\end{equation}
where $|b|<1$ and $\varepsilon_{it}\sim N(0,1-b^2)$. The FRV techniques provide analytic derivation for the eigenvalue distribution of ${\Sigma}_{model}^{(b)}=\frac{1}{T}UU^{T}$. The implementation steps are briefly described here.
\begin{enumerate}[1.]
\item Get $\rho_{model}(\lambda;b)$ from the Green's function $G_{\Sigma_{model}}(z)$:
\begin{equation}
\label{Eq:inverse_green_function}
\begin{aligned}
  \rho_{model}(\lambda;b) = -\frac{1}{\pi} \lim\limits_{\epsilon\rightarrow 0^{+}}\Im G_{\Sigma_{model}}(\lambda + i\epsilon)
\end{aligned},
\end{equation}
where $\Im$ represents getting the imaginary part operation, $\lambda$ is the eigenvalue variable and $\epsilon$ is the imaginary part.
\item The Green's function $G_{\Sigma_{model}}(z)$ can be obtained from the moments' generating function $M_{\Sigma_{model}}(z)$:
\begin{equation}
\label{Eq:green_function}
\begin{aligned}
  G_{\Sigma_{model}}(z) = \frac{M_{\Sigma_{model}}(z)+1}{z} \qquad for\; |z|\neq 0
\end{aligned}.
\end{equation}
\item Solve the polynomial equation for $M \equiv M_{\Sigma_{model}}(z)$:
\begin{equation}
\label{Eq:polynomial}
\begin{aligned}
  a^4c^2M^4+2a^2c(-(1+b^2)z+a^2c)M^3+((1-b^2)^2z^2 \\
  -2a^2c(1+b^2)z+(c^2-1)a^4)M^2-2a^4M-a^4=0
\end{aligned},
\end{equation}
where $a = \sqrt{1-b^2}$ and $c = \frac{N}{T}$. In practice, the $4$th order polynomial can be solved by using $numpy.roots()$ function in $Python$. For the multiple roots obtained, the largest one will be selected.
\end{enumerate}
See Appendix \ref{section: derivation} for details.

The spectral distance measure $\mathcal{D}$ must be sensitive to the information disparity in $\rho_{real}(p)$ and $\rho_{model}(b)$. Here, we use Jensen-Shannon divergence, a symmetrized version of Kullback-Leibler divergence, which is defined as
\begin{equation}
\label{Eq:spectral_distance}
\begin{aligned}
  \mathcal{D}({\rho_{real}}||{\rho_{model}}) = \frac{1}{2}\sum\limits_{i}{\rho_{real}^{(i)}}\log {\frac{{\rho_{real}^{(i)}}}{{\rho}^{(i)}}} \\
  +\frac{1}{2}\sum\limits_{i}{\rho_{model}^{(i)}}\log {\frac{\rho_{model}^{(i)}}{{\rho}^{(i)}}}
\end{aligned},
\end{equation}
where ${\rho} = \frac{{\rho_{real}}+{\rho_{model}}}{2}$. It is noted that $\mathcal{D}({\rho_{real}}||{\rho_{model}})$ becomes smaller as $\rho_{real}$ approaches $\rho_{model}$, and vice versa. Therefore, the optimal parameter set $({\hat p},{\hat b})$ can be obtained by minimizing the spectral distance $\mathcal{D}$.

\subsection{Spatio-Temporal Correlation Analysis Approach for Anomaly Detection and Location}
\label{section: approach}
From Section \ref{section: problem}, we know that the number of removed factors $p$ and the autoregressive rate $b$ can be used to indicate the variations of spatial and temporal correlation of the data. Based on the estimated parameter $\hat p$, we design a partial linear eigenvalue statistics for the eigenvalues corresponding to the removed $\hat p$ factors to measure the spatial correlation, which is defined as
\begin{equation}
\label{Eq:les_partial}
\begin{aligned}
  \mathcal{N}_{\phi} = \sum\limits_{i = 1}^{\hat p} {\phi ({\lambda _{i}})}
\end{aligned},
\end{equation}
where $\lambda_1>\lambda_2>\cdots>\lambda_{\hat p}$, and $\phi(\cdot)$ is a test function that makes a linear or nonlinear mapping for the eigenvalues $\lambda_i$. The commonly used test functions include chebyshev polynomial (such as $\phi(\lambda) = 2{\lambda}^2-1$), information entropy (i.e., $\phi(\lambda) = -\lambda ln\lambda$), likelihood radio function (i.e., $\phi(\lambda) = \lambda-ln\lambda-1$), and wasserstein distance (i.e., $\phi(\lambda) = \lambda-2\sqrt\lambda+1$). More details about the test functions can be found in our previous work \cite{shi2018incipient}. As an indicator to measure the spatial correlation of the data, $\mathcal{N}_{\phi}$ is more accurate and robust than the estimated number of factors $\hat p$, because the latter is susceptible to the weak factors caused by random fluctuations. Meanwhile, the estimated parameter $\hat b$ is directly used to measure the temporal correlation of the real data. It can effectively emulate the variation of the temporal correlation of the data, and provide an insight into system dynamics. To be mentioned is that, if the residual processes of the real data are not auto-correlated, $\hat b$ will be far different from the true value.

According to the matrix theory, the contribution rate of the $j-$th ($1\le j\le N$) row to the eigenvalue $\lambda_i$ of a covariance matrix can be measured by the $j-$th element of the corresponding principal component ${\hat F}^{(i)}$. See Appendix \ref{section: proof} for proofs. This inspires us to realize anomaly location by using the estimated $\hat p$ factors and the corresponding eigenvectors. An anomaly location indicator is designed as
\begin{equation}
\label{Eq:location_indicator}
\begin{aligned}
  \bm\eta = \sum\limits_{i = 1}^{\hat p} {{\lambda _{i}}{|{\hat F}^{(i)}|}}
\end{aligned},
\end{equation}
where $\bm\eta$ is a vector of length $N$.

In real applications, we can move a certain length window on the collected data set ${\bf D}$ at continuous sampling times and the last sampling time is the current time, which enables us to track the variations of spatio-temporal correlations of the online monitoring data in real-time. For example, at the sampling time $t_j$, the obtained raw data matrix ${\bf R}(t_j)\in{\mathbb{R}^{N\times T}}$ is formulated by
\begin{equation}
\label{Eq:matrix_formulate}
\begin{aligned}
  {\bf{R}}(t_j) = \left( {{\bf{d}}(t_{j-T+1}),{\bf{d}}(t_{j-T+2}), \cdots ,{\bf{d}}(t_j)} \right)
\end{aligned},
\end{equation}
where ${\bf d}(t_k)={({d_1,d_2,\cdots,d_N})}^H$ for $t_{j-T+1}\le t_k \le t_j$ is the sampling data at time $t_k$. Thus, $\mathcal{N}_{\phi}(t_j)$, $\hat b(t_j)$ and $\bm\eta(t_j)$ are produced for the sampling time $t_j$. In order to realize anomaly declare automatically, the confidence level $1-\alpha$ of each anomaly indicator is calculated and compared with the defined threshold $(1-\alpha)_{th}$. Take $\mathcal{N}_{\phi}$ for example, for a series of time $T'\;(t_{j-T'+1}\sim t_j)$, $\mathcal{N}_{\phi}$ is considered to follow a student's t-distribution with $T'-1$ degrees of freedom. At the sampling time $t_j$, the anomaly indicator $\mathcal{N}_\phi(t_j)$ is standardized by
\begin{equation}
\label{Eq:threshold}
\begin{aligned}
  {\hat{\mathcal{N}_{\phi}}(t_j)} = \frac{{\mathcal{N}_{\phi}(t_j)}-\mu ({\mathcal{N}_{\phi}})}{\sigma ({\mathcal{N}_{\phi}})}
\end{aligned},
\end{equation}
where ${\mathcal{N}_{\phi}(t_j)}\in {\mathcal{N}_{\phi}}$, $\mu ({\mathcal{N}_{\phi}})$ and $\sigma ({\mathcal{N}_{\phi}})$ are the mean and standard deviation of $ {\mathcal{N}_{\phi}}$, and $\hat{\mathcal{N}_{\phi}}$ follows the standard t-distribution. Thus, We can obtain the confidence level $1-\alpha$ of $\mathcal{N}_\phi (t_j)$ once ${\hat{\mathcal{N}_{\phi}}(t_j)}$ is calculated. For example, let ${\hat{\mathcal{N}_{\phi}}(t_j)}=2.650$ and $T'=14$, then the confidence level $1-\alpha$ is $98\%$. Thus, the anomaly can be declared automatically by comparing $1-\alpha$ with $(1-\alpha)_{th}$, .

Based on the research mentioned above, an anomaly detection and location approach based on spatio-temporal correlation analysis is designed. The fundamental steps are given as follows. Steps $4\sim 8$ are conducted for calculating the ESD of the covariance matrix of the real residuals, Steps $9\sim 10$ are for calculating the limiting spectral density of the built covariance model, and the spectral distance of them are calculated and saved in each iteration shown in Step $11$. In Step $12$, the optimal parameter set corresponding to the minimum spectral distance is obtained for each sampling time. Based on the steps above, $\mathcal{N}_{\phi}$ and $\hat b$ are calculated as indicators to detect anomalies and $\bm\eta$ is calculated for anomaly location.
\begin{table}[htbp]
\label{Tab: algorithm}
\centering
\footnotesize
\begin{tabular}{p{8.4cm}}   
\toprule[1.0pt]
\textbf {Steps of spatio-temporal correlation analysis for anomaly detection and location in distribution networks}\\
\hline
1. For each feeder, construct a spatio-temporal data set $\bf D$ by arranging \\
\quad three-phase voltage measurements from all monitoring devices within \\
\quad the feeder in chronological order.  \\
2. At each sampling time $t_j$:  \\
3.\quad Obtain the corresponding data matrix ${\bf R}(t_j)\in\mathbb{R}^{N\times T}$ by using \\
\quad\quad an $N\times T$ window on $\bf D$; \\
4.\quad For the number of removing factors $p = 1,2,\cdots$ \\
5.\quad\quad Get the real residuals ${\hat U}^{(p)}(t_j)$ through equation (\ref{Eq:residual}); \\
6.\quad\quad Normalize ${\hat U}^{(p)}(t_j)$ into the standard form through equation (\ref{Eq:standardize}); \\
7.\quad\quad Calculate the covariance matrix of the standardized ${\hat U}^{(p)}(t_j)$, i.e., \\
\quad\quad\; ${\Sigma}_{real}^{(p)}(t_j)$; \\
8.\quad\quad Obtain the ESD of ${\Sigma}_{real}^{(p)}(t_j)$, i.e., $\rho_{real}^{(p)}(t_j)$; \\
9.\quad\quad For the autoregressive rate $b\sim U[0,1]$ \\
10.\quad\quad\quad Obtain $\rho_{model}^{(b)}(t_j)$ through equation (\ref{Eq:inverse_green_function}),  (\ref{Eq:green_function}) and (\ref{Eq:polynomial}); \\
11.\quad\quad\quad Calculate the spectral distance $\mathcal{D}({\rho_{real}^{(p)}(t_j)}||{\rho_{model}^{(b)}(t_j)})$ \\ \quad\quad\quad\quad\; through equation (\ref{Eq:spectral_distance}) and save them; \\
12.\; Obtain the optimal parameter set $({\hat p}(t_j), {\hat b}(t_j))$ through equation (\ref{Eq:distance}); \\
13.\; Calculate the spatial indicator $\mathcal{N}_{\phi}(t_j)$ through equation (\ref{Eq:les_partial}); \\
14.\; Calculate the location indicator $\bm\eta(t_j)$ through equation (\ref{Eq:location_indicator}); \\
15. Draw the $\mathcal{N}_{\phi}-t$, ${\hat b}-t$ and $\bm\eta-t$ curves for each feeder in a series \\
\quad\; of time to realize anomaly detection and location. \\
\hline
\end{tabular}
\end{table}

The anomaly detection approach proposed is driven by the online monitoring data in distribution networks, and based on high-dimensional statistical theories. It reveals the variations of spatio-temporal correlations of the input data when anomalies occur and can detect the anomalies in an early phase by controlling both the number of factors and the autoregressive rate. Compared with traditional model-based methods, the proposed approach is purely driven by data and does not require too much prior knowledge about the complex topology of the distribution network. It is robust against small random fluctuations and measuring errors in the network, which can help reduce the false alarming rate. What's more, the proposed approach is practical for real-time anomaly detection and location by moving a certain length window method.
\subsection{More Discussions About the Proposed Approach}
\label{section: more_disscussion}
\begin{figure}[htb]
\centering
\begin{minipage}{4.1cm}
\centerline{
\includegraphics[width=1.9in]{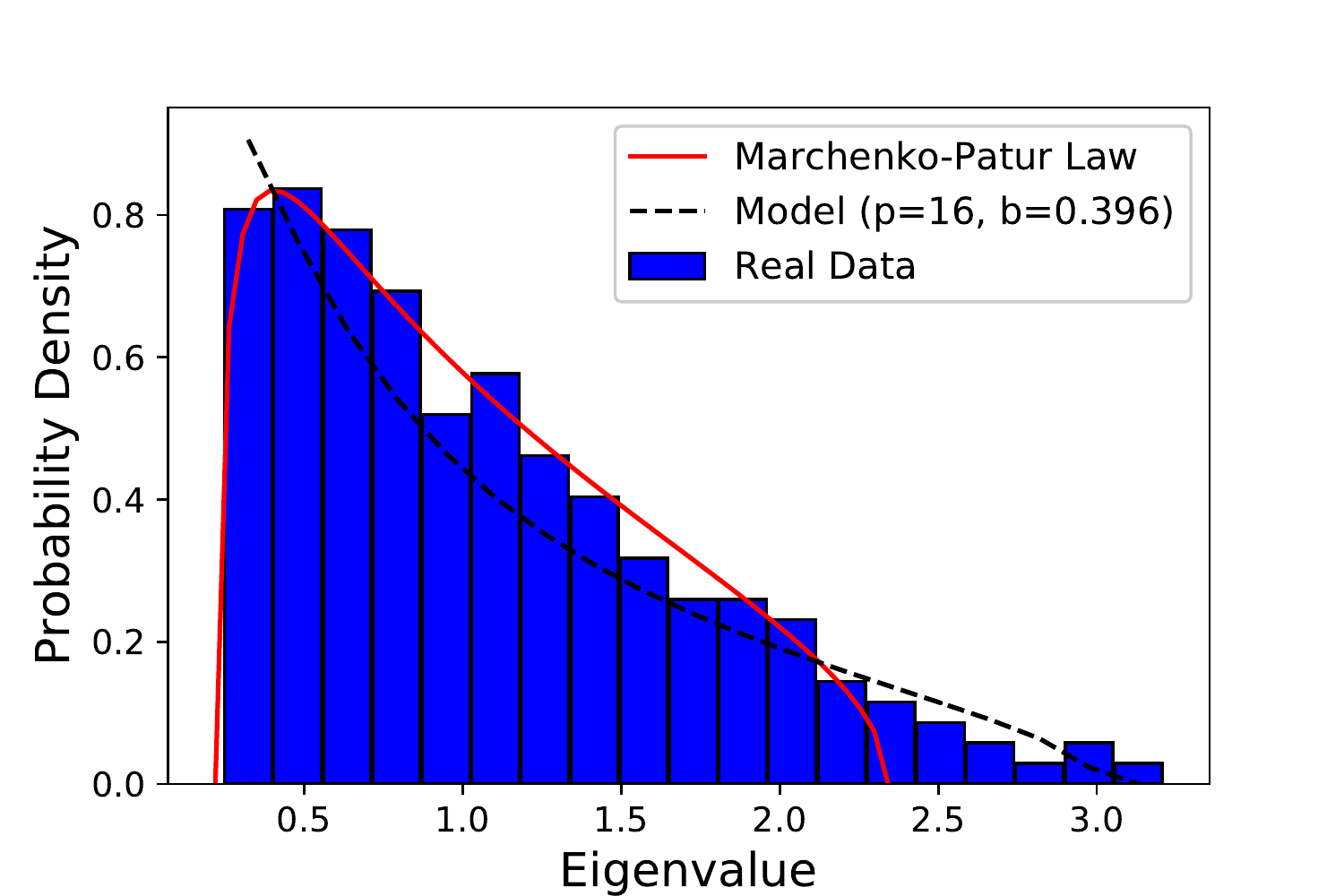}
}
\parbox{5cm}{\small \hspace{1.2cm}(a) Normal state }
\end{minipage}
\hspace{0.2cm}
\begin{minipage}{4.1cm}
\centerline{
\includegraphics[width=1.9in]{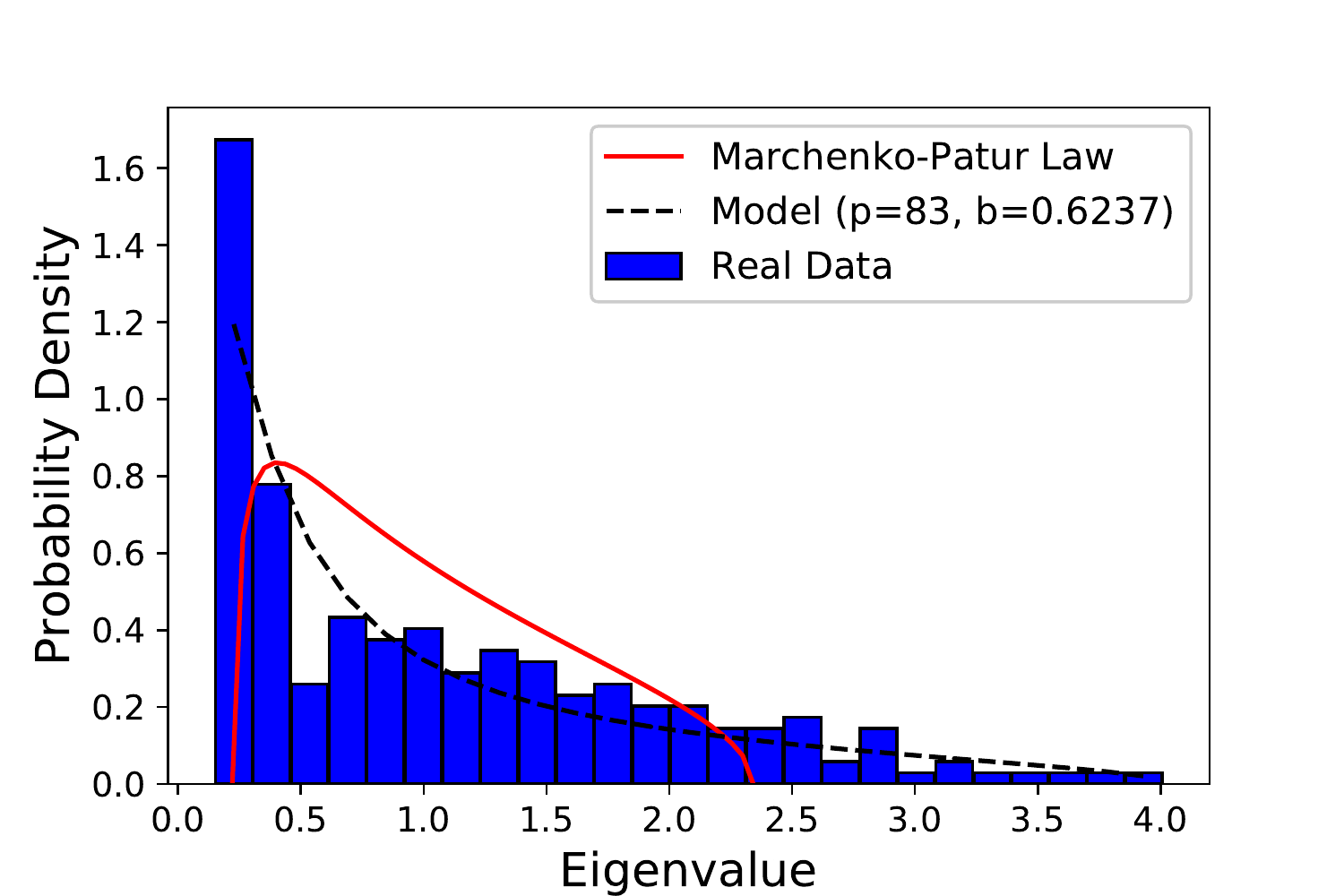}
}
\parbox{5cm}{\small \hspace{1.2cm}(b) Abnormal state }
\end{minipage}
\caption{Fit of our built model to the real residuals constructed from the real-world online monitoring data. The built model with estimated $\hat p$ and $\hat b$ fits the real residuals very well (the spectral distance between the model and residuals: 0.026 (left) and 0.028 (right)). For comparison, the M-P law for the real residuals is plotted (the spectral distance between the M-P law and residuals: 0.109 (left) and 0.296 (right)).}
\label{fig:residual_example}
\end{figure}

The first issue we want to discuss is the assumptions made in the proposed approach. In Section \ref{section: residual}, we assume that the cross-correlations of the real residuals can be effectively eliminated by removing $p$ factors and the temporal correlations of them are exponentially decreasing. However, for the real-world online monitoring data in a distribution network, whether this assumption holds is questionable. Meanwhile, the factor model estimation method in Section \ref{section: factor_model_estimation} is suitable for large-dimensional data matrix in theory. However, in practice, the dimensions of the online monitoring data for some feeder lines are moderate, such as hundreds or less. Here, we will check how well our built covariance model can fit the real residuals, results of which are shown in Figure \ref{fig:residual_example}.

Figure \ref{fig:residual_example}(a) and \ref{fig:residual_example}(b) respectively show the fitting result of our built covariance model to the real residuals under both normal and abnormal feeder line operating states. It can be observed that, with optimal parameter set $({\hat p}, {\hat b})$, our built model can fit the real residuals well no matter whether the feeder line operates in normal or abnormal state. In contrast, the M-P law does not fit the real residuals. The well fitted result validates our assumption for the real residuals, and it verifies the feasibility of the proposed approach for analyzing the medium dimensional data. Furthermore, it is noted that the estimated $\hat p$ and $\hat b$ are different when the feeder line operates in different states, which explains why they can be used as basic indicators to detect the anomalies.

The second issue we want to discuss here is how can the proposed approach be integrated into distribution management system (DMS). In Section \ref{section: approach}, we calculate the confidence level $1-\alpha$ of the anomaly indicator for each sampling time and compare it with the threshold $(1-\alpha)_{th}$ for declaring an anomaly. In practice, we can divide the operating states of the feeders into emergency, high risk, preventive and normal, and combine them with the calculated values of $1-\alpha$. For example, if $1-\alpha>90\%$, the operating state of the feeder is diagnosed as in emergency state and further analysis will be conducted. In this way, the proposed approach can be used for assessing the operational risks of feeders in DMS.

Another issue is the delay tolerance. The data collected from different monitoring devices will arrive with different delays, which will cause data disalignment or incompletion. In theory, the proposed approach is a correlation analysis approach based on spectrum analysis, which has been proved to be robust to the data disalignment in \cite{he2018novel}. In practice, compared with the large size data window for each sampling time, the data disalignment caused by small delay can almost be ignored. If high data delay exists, the data collected can be divided into different groups according to the delay tolerance. The data matrix formulated in each group is analyzed by the proposed approach and the results are fused to serve as the anomaly indicator.
\section{Case Studies}
\label{section: case_study}
In this section, the proposed anomaly detection and location approach is validated with both synthetic data from IEEE 33-bus and 57-bus test systems and the real-world online monitoring data in a distribution network. Detailed information about IEEE 33-bus and 57-bus test systems can be found in case33.m and case57.m in Matpower package \cite{5491276}. The simulation environment is MATLAB2016. Five cases in different scenarios were designed: 1) The first case, leveraging the synthetic data from IEEE 33-bus distribution test system, tested the effectiveness of the proposed approach for anomaly detection and location. 2) The implications of parameter $p$ and $b$ involved in the approach were explored in the second case. The synthetic data was generated from IEEE 57-bus test system which can be considered as a distribution network system connected with distributed generators; 3) In the third case, we illustrated the advantages of the proposed approach in anomaly detection by comparing it with other existing techniques. 4) The last two cases, using the real-world online monitoring data, validated the effectiveness and advantages of the proposed approach.
\subsection{Case Study with Synthetic Data}
\label{section: case_synthetic}
\begin{figure}[!t]
\centerline{
\includegraphics[width=3.0in]{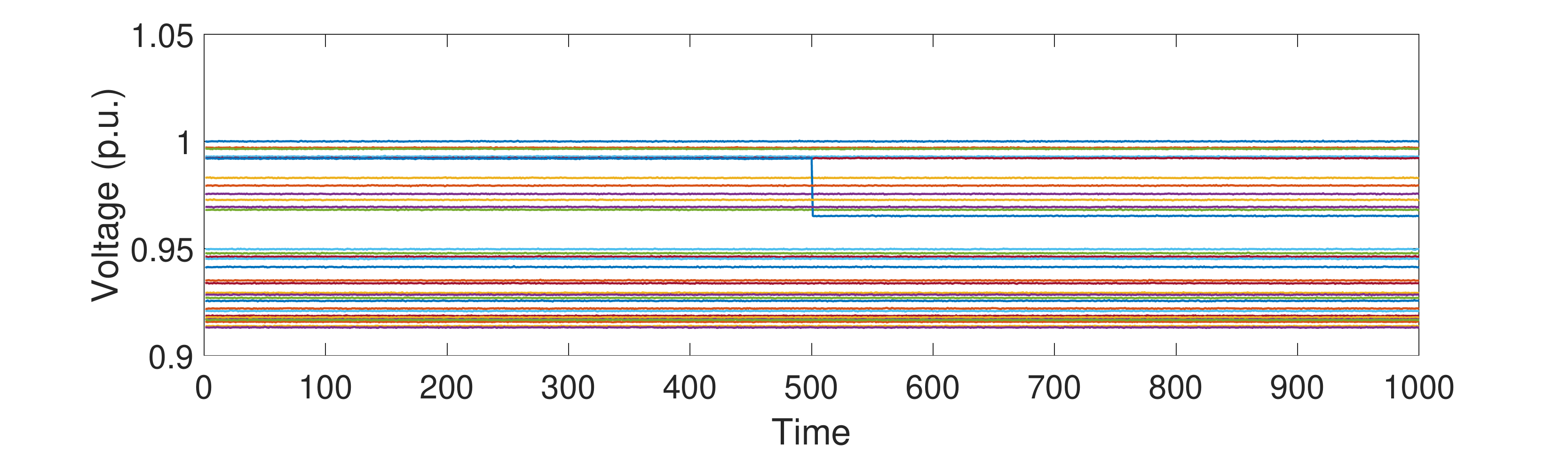}
}
\caption{The synthetic data generated from IEEE 33-bus distribution test system. One anomaly signal was set at $t_s=501$.}
\label{fig:case1_org}
\end{figure}

1) Case Study on the Effectiveness of the Proposed Approach: In this case, the synthetic data generated from IEEE 33-bus distribution test system contained $33$ voltage measurement variables with sampling $1000$ times. In order to test the effectiveness of the proposed approach, an assumed anomaly signal was set by a sudden increase of impedance from bus $21$ to $22$ and others stayed unchanged, which was shown in Table \ref{Tab: Case1}. The generated data is shown in Figure \ref{fig:case1_org}. In the experiment, the size of the moving window was set to be $33\times 200$. For each moving window $\bf D$, the autoregressive (AR) noise with a decaying rate $b=0.5$ (i.e., $E_{it}=0.5*E_{i,t-1}+\varepsilon_{it}$, where $\varepsilon_{it}\sim N(0,1-0.5^2)$ so that the variance of $E_t$ is $1$.) was introduced into the data to represent random fluctuations and measuring errors. The scale of the added AR noise is calculated as $m=\sqrt{\frac{var (\bf D)}{var {(\bf E)}*SNR}}$, where $var (\cdot)$ denotes the variance operation, and $SNR$ is the signal-noise-rate which was set to be $500$. The experiment was repeated for 20 times and the results were averaged. Here, we chose the likely-hood radio function (i.e., $\phi (\lambda)={\lambda}-ln{\lambda}-1$;) as the test function in equation (\ref{Eq:les_partial}).
\begin{table}[!t]
\caption{Assumed Signals From Bus $21$ to $22$ in Case 1.}
\label{Tab: Case1}
\centering
\footnotesize
\begin{tabular}{cclc}   
\toprule[1.0pt]

\textbf {fBus} & \textbf {tBus} & \textbf{Sampling Time}& \textbf{Impedance(p.u.)}\\
\hline
\multirow{2}*{21} & \multirow{2}*{22} & $t_s=1\sim 500$ & 0.5 \\
~ & ~ & $t_s=501\sim 1000$ & 20 \\
Others & Others & $t_s=1\sim 1000$ & Unchanged \\
\hline
\end{tabular}
\end{table}
\begin{figure}[!t]
\centering
\begin{minipage}{4.1cm}
\centerline{
\includegraphics[width=1.9in]{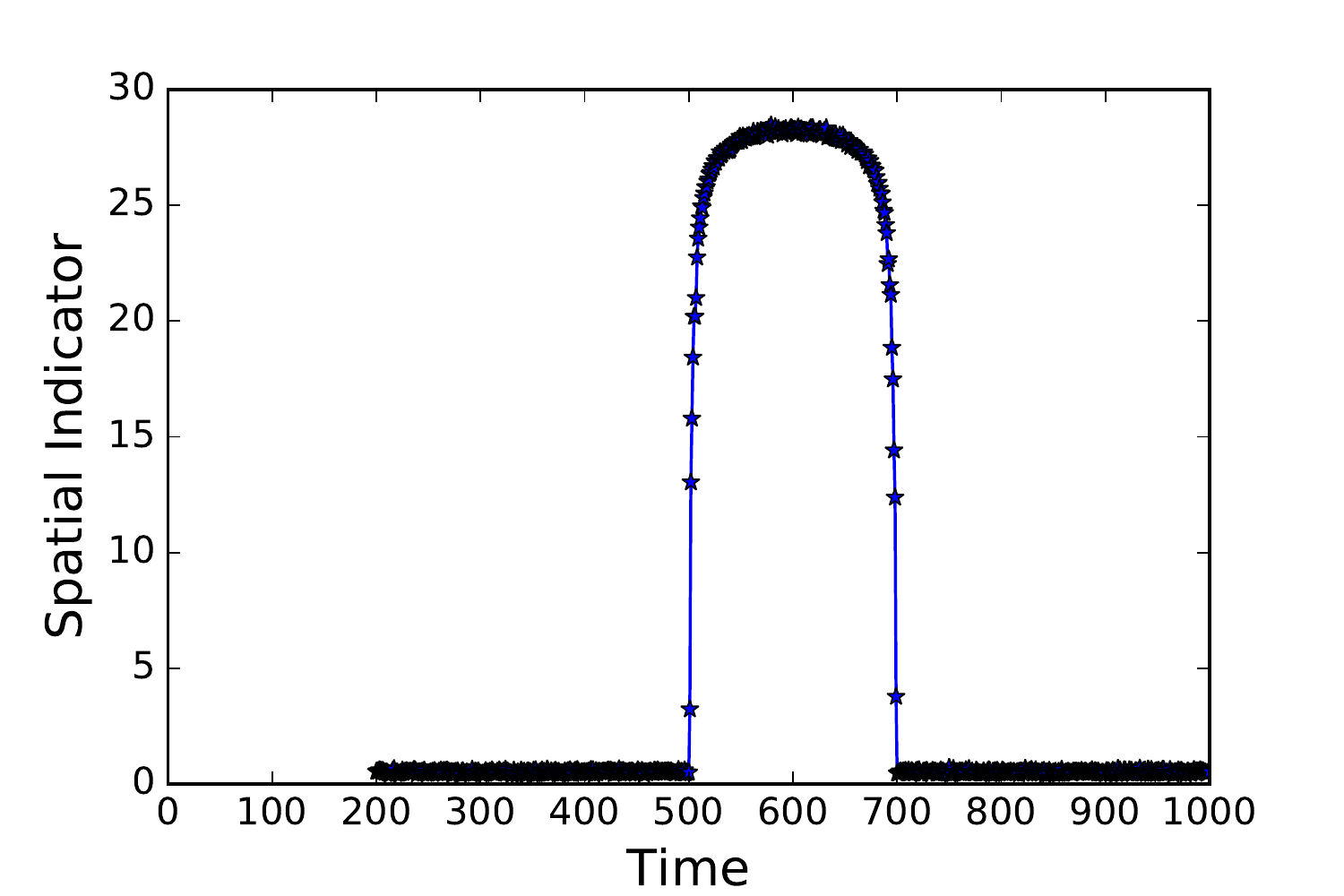}
}
\parbox{5cm}{\small \hspace{1.2cm}(a) $\mathcal{N}_{\phi}-t$ curve}
\end{minipage}
\hspace{0.2cm}
\begin{minipage}{4.1cm}
\centerline{
\includegraphics[width=1.9in]{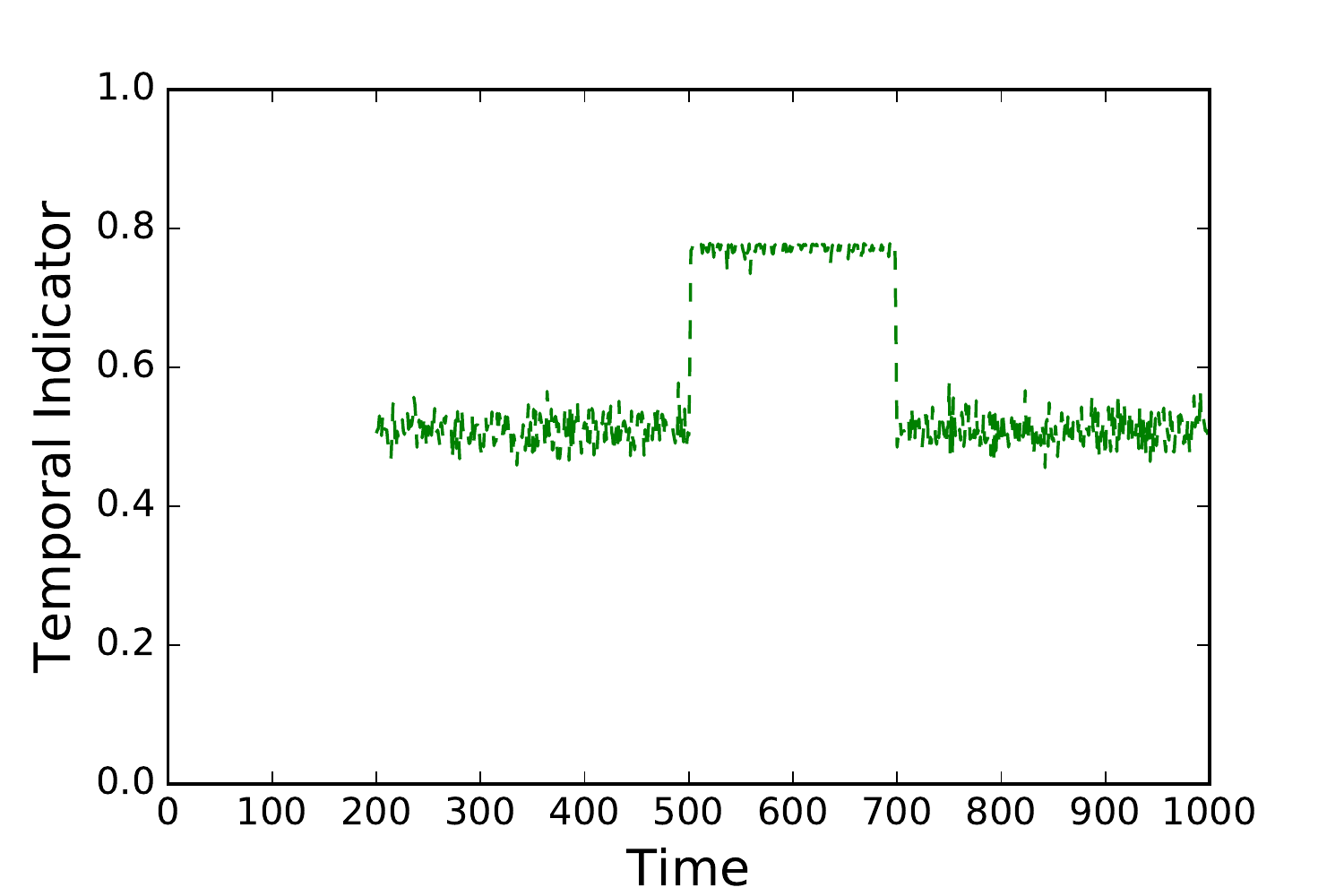}
}
\parbox{5cm}{\small \hspace{1.2cm}(b) ${\hat b}-t$ curve}
\end{minipage}
\hspace{0.2cm}
\caption{The anomaly detection result in Case 1. }
\label{fig:case1_indicator}
\end{figure}
\begin{figure}[!t]
\centerline{
\includegraphics[width=3.0in]{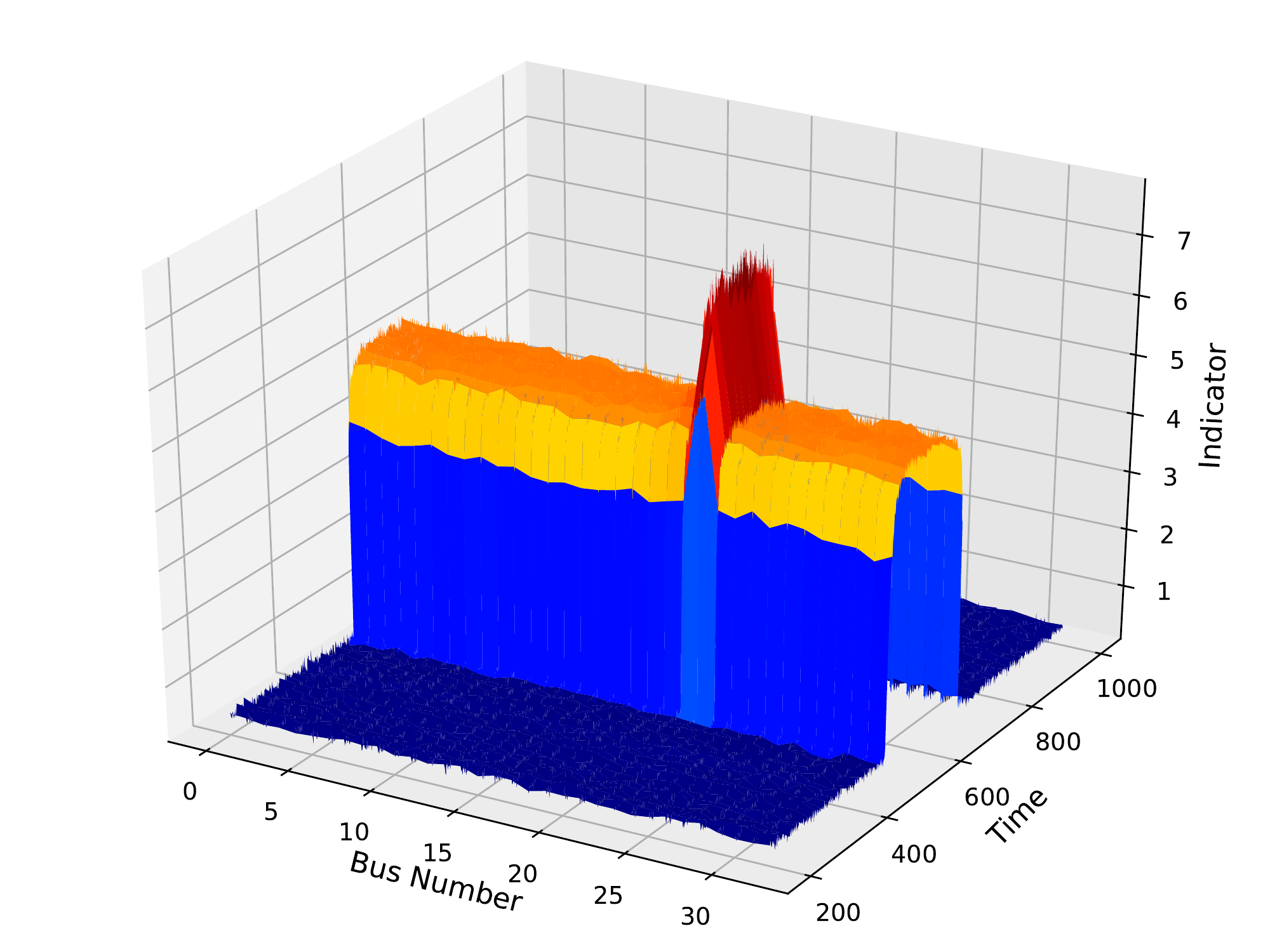}
}
\caption{The anomaly location result in Case 1.}
\label{fig:case1_loc}
\end{figure}

Figure \ref{fig:case1_indicator}(a) and Figure \ref{fig:case1_indicator}(b) show the $\mathcal{N}_{\phi}-t$ and ${\hat b}-t$ curves generated with continuously moving windows. It is noted that the curves begin at $t_s=200$, because the initial window includes 199 times of historical sampling and the present sampling data. In calculating the confidence level $1-\alpha$ for each data point in the detection curves, $\mathcal{N}_{\phi}$ and ${\hat b}$ during continuous $200$ points (199 historical points and the current point) were considered to follow the student's t-distribution. The detection processes are shown as follows:

\uppercase\expandafter{\romannumeral1}. During $t_s=200\sim 500$, $\mathcal{N}_{\phi}$ and $\hat b$ remain almost constant and the corresponding values of $1-\alpha$ are small, which means the system operates in normal state and the spatio-temporal correlations of the data stay almost unchanged. For example, at $t_s=500$, the calculated $1-\alpha$ of $\mathcal{N}_{\phi}$, $\hat b$ are $34.123\%$, $29.294\%$, respectively.

\uppercase\expandafter{\romannumeral2}. From $t_s=501$, $\mathcal{N}_{\phi}$ and $\hat b$ begin to change and the corresponding values of $1-\alpha$ increase rapidly, which indicates an anomaly signal occurs and the spatio-temporal correlations of the data begin to change. For example, at $t_s=501$, the calculated $1-\alpha$ of $\mathcal{N}_{\phi}$, $\hat b$ are $99.328\%$, $99.999\%$, respectively. It is noted that $\mathcal{N}_{\phi}-t$ and $\hat b-t$ curves are almost inverted U-shape, because the delay lag of the anomaly signal to the spatio-temporal indicators is equal to the window's width.

Furthermore, the anomaly is located through the proposed approach, result of which is shown in Figure \ref{fig:case1_loc}. It can be observed that, from $t_s=501$, the location indicator $\bm\eta$ increases rapidly and $\eta_{21}$ is significantly higher than others, which indicates anomaly occurred on bus $21$. For example, at $t_s=501$, the calculated $1-\alpha$ corresponding to bus $21$ and others (such as bus $20$) are $99.682\%$ and $21.194\%$, respectively. The anomaly location result coincides with the assumed signal location in Table \ref{Tab: Case1}.

2) Case Study on the Implications of $p$ and $b$: In case 1, it is observed the estimated $\hat p$ and $\hat b$ are different when the system operates in different states. In this case, we will further explore what drives them. The IEEE 57-bus test system can be considered as a distribution network connected with distributed generators, and it was used to generate the synthetic data. During the simulations, a change of the active load at one bus was considered as an anomaly event.

\begin{table}[!t]
\caption{Assumed Signals for Active Load of Bus 20, 30 and 40 in Case 2.}
\label{Tab: p_Case2}
\centering
\footnotesize
\begin{tabular}{clc}   
\toprule[1.0pt]
\textbf {Bus} & \textbf{Sampling Time}& \textbf{Active Power(MW)}\\
\hline
\multirow{2}*{20} & $t_s=1\sim 500$ & 5 \\
~&$t_s=501\sim 1000$ & 10 \\
\multirow{2}*{30} & $t_s=1\sim 510$ & 5 \\
~&$t_s=511\sim 1000$ & 10 \\
\multirow{2}*{40} & $t_s=1\sim 520$ & 5 \\
~&$t_s=521\sim 1000$ & 10 \\
Others & $t_s=1\sim 1000$ & Unchanged \\
\hline
\end{tabular}
\end{table}
\begin{figure}[!t]
\centerline{
\includegraphics[width=3.0in]{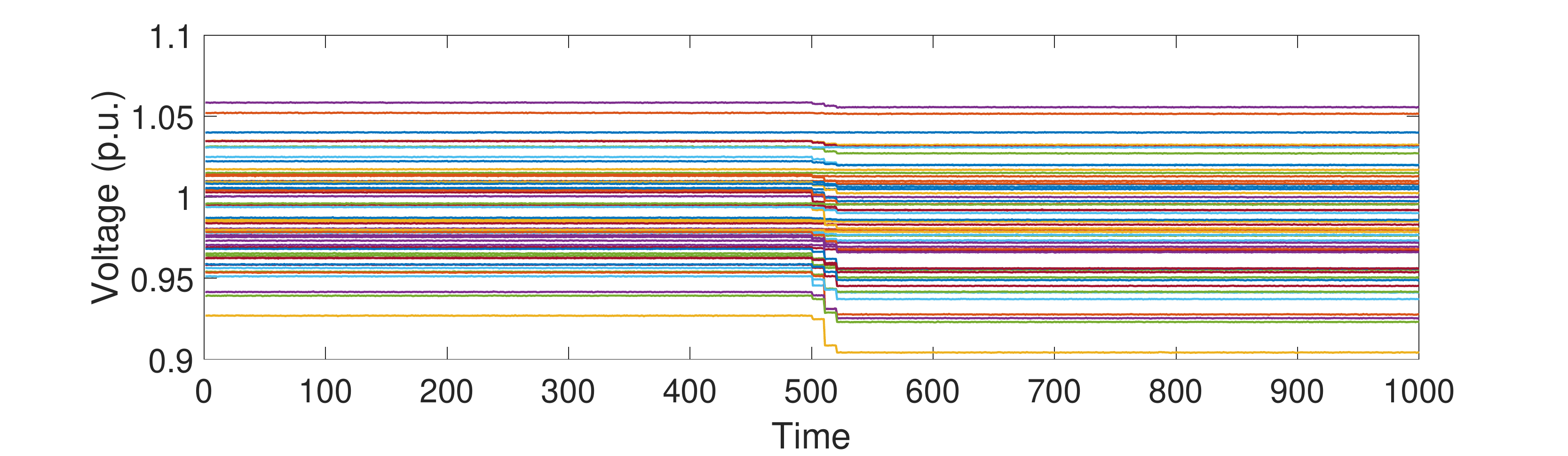}
}
\caption{The synthetic data generated from IEEE 57-bus test system. Multiple anomaly signals were set at $t_s=501$, $t_s=511$, $t_s=521$, respectively.}
\label{fig:p_case2_org}
\end{figure}
\begin{figure}[!t]
\centering
\begin{minipage}{4.1cm}
\centerline{
\includegraphics[width=1.9in]{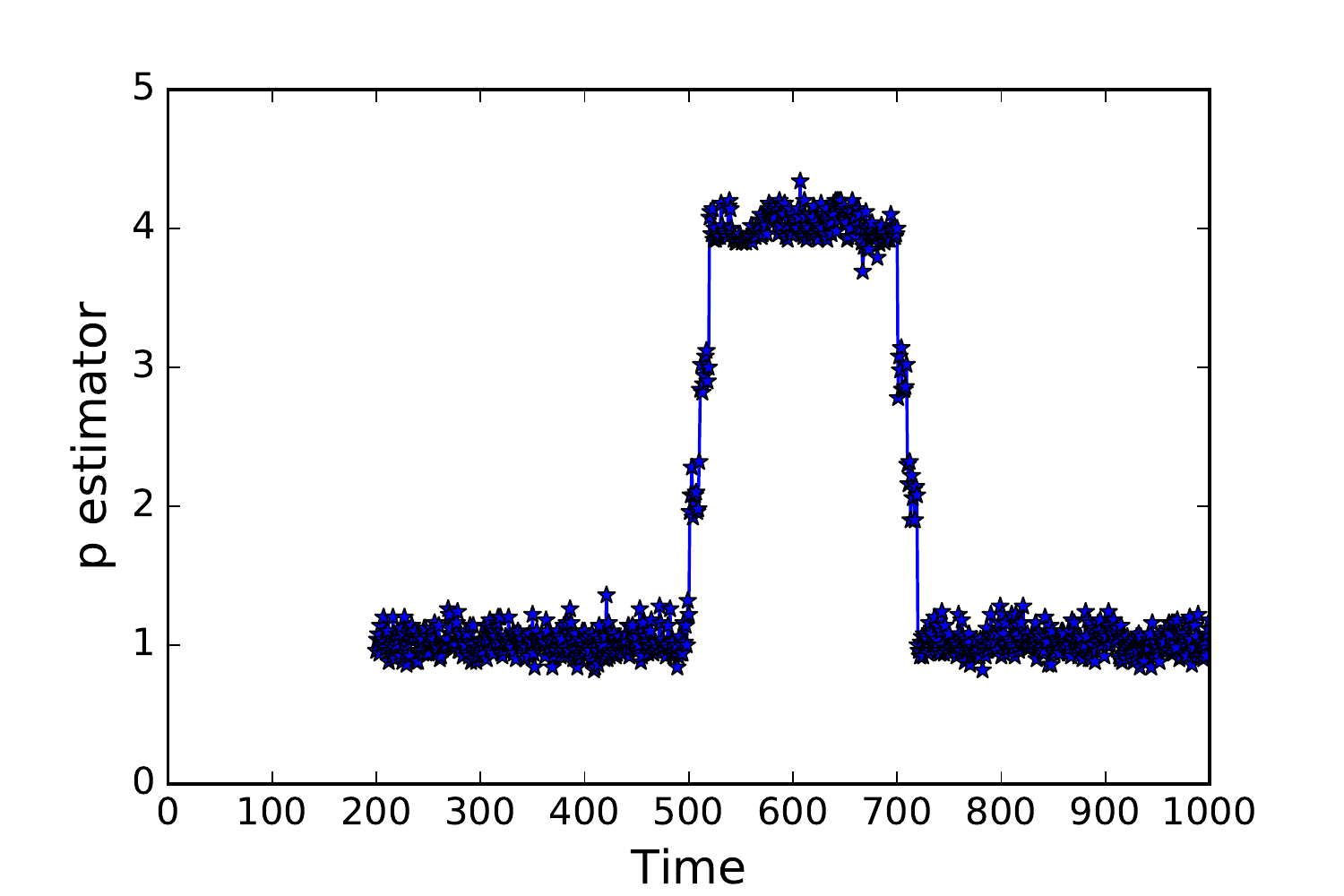}
}
\parbox{5cm}{\small \hspace{1.2cm}(a) ${\hat p}-t$ curve}
\end{minipage}
\hspace{0.2cm}
\begin{minipage}{4.1cm}
\centerline{
\includegraphics[width=1.9in]{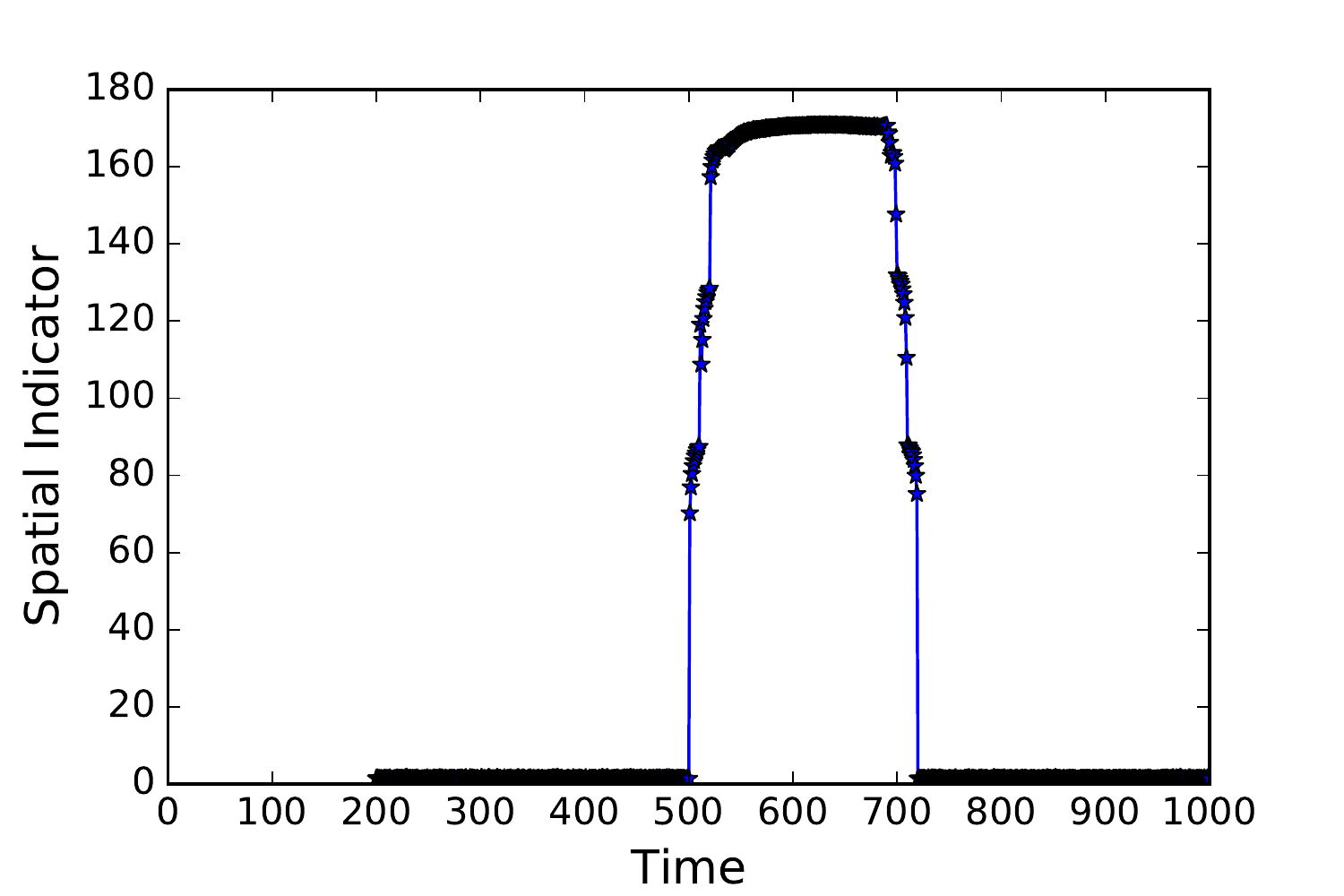}
}
\parbox{5cm}{\small \hspace{1.2cm}(b) $\mathcal{N}_{\phi}-t$ curve}
\end{minipage}
\caption{Multiple anomaly signal detection result in Case 2.}
\label{fig:p_case2_indicator}
\end{figure}
In order to interpret $p$, multiple anomaly signals were set, which is shown in Table \ref{Tab: p_Case2}. The generated data is shown in Figure \ref{fig:p_case2_org}. In the experiment, the size of the moving window was set to be $57\times 200$ and the other parameters were set the same as in case 1). The experiment was repeated for $20$ times with results being averaged. The generated ${\hat p}-t$ curve and $\mathcal{N}_{\phi}-t$ curve with continuously moving windows are shown in Figure \ref{fig:p_case2_indicator}. Interpretations of $p$ are stated as follows:

\uppercase\expandafter{\romannumeral1}. During $t_s=200\sim 500$, $\hat p$ and $\mathcal{N}_{\phi}$ remain nearly $1$ and $1.37$, which means no strong factor appears.

\uppercase\expandafter{\romannumeral2}. From $t_s=500$ to $t_s=501$, $\hat p$ and $\mathcal{N}_{\phi}$ increase from nearly $1$, $1.37$ to $2$, $70.23$, respectively, which indicates one strong factor is estimated. From $t_s=510$ to $t_s=511$, $\hat p$ and $\mathcal{N}_{\phi}$ increase from nearly $2$, $87.46$ to $3$, $119.09$, respectively, which indicates another new strong factor is estimated. Similar analysis result can be obtained from $t_s=520$ to $t_s=521$. Combining the anomaly signals set in Table \ref{Tab: p_Case2}, it can be concluded that $\hat p$ is driven by the number of anomaly events.

\uppercase\expandafter{\romannumeral3}. From $t_s=701$ to $t_s=730$, $\hat p$ decreases by $1$ per $10$ sampling times, which coincides with the decrease of the number of anomaly signals contained in the moving window.

\begin{table}[!t]
\caption{An Assumed Signal for Active Load of Bus 20 in Case 2.}
\label{Tab: b_Case2}
\centering
\footnotesize
\begin{tabular}{clc}   
\toprule[1.0pt]
\textbf {Bus} & \textbf{Sampling Time}& \textbf{Active Power(MW)}\\
\hline
\multirow{2}*{20} & $t_s=1\sim 500$ & $10$ \\
~&$t_s=501\sim 1000$ & $10\rightarrow 60$ \\
Others & $t_s=1\sim 1000$ & Unchanged \\
\hline
\end{tabular}
\end{table}
\begin{figure}[!t]
\centerline{
\includegraphics[width=3.0in]{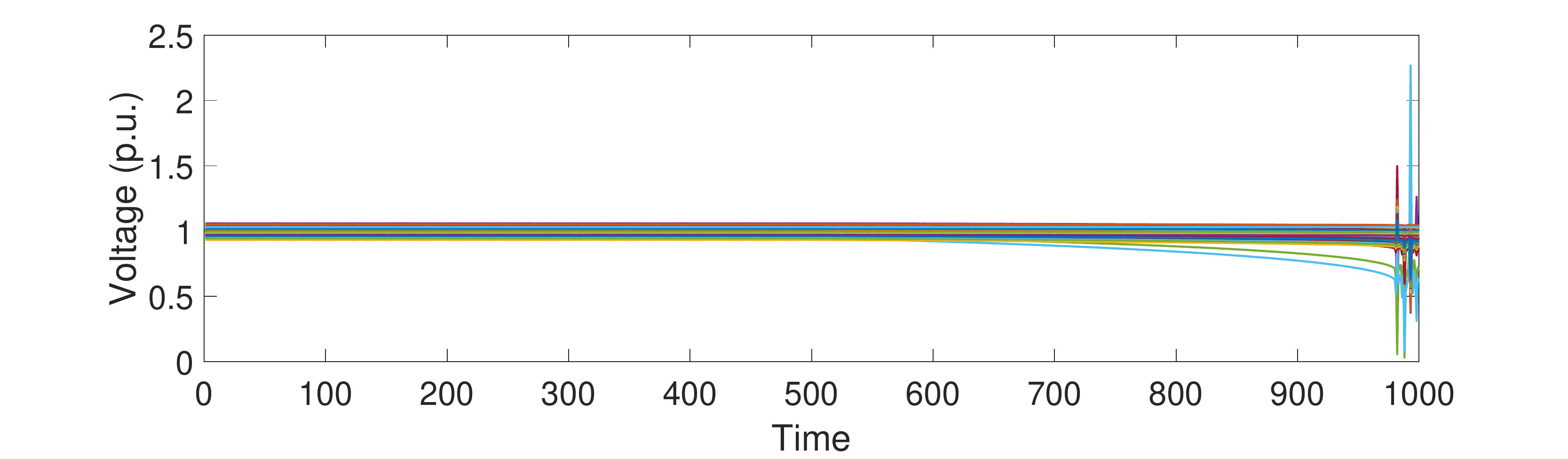}
}
\caption{The synthetic data generated from IEEE 57-bus test system. An increasing anomaly signal was set at $t_s=501$. With the increase of the anomaly signal, the voltage collapses at $t_s=980$.}
\label{fig:b_case2_org}
\end{figure}
\begin{figure}[!t]
\centering
\begin{minipage}{4.1cm}
\centerline{
\includegraphics[width=1.9in]{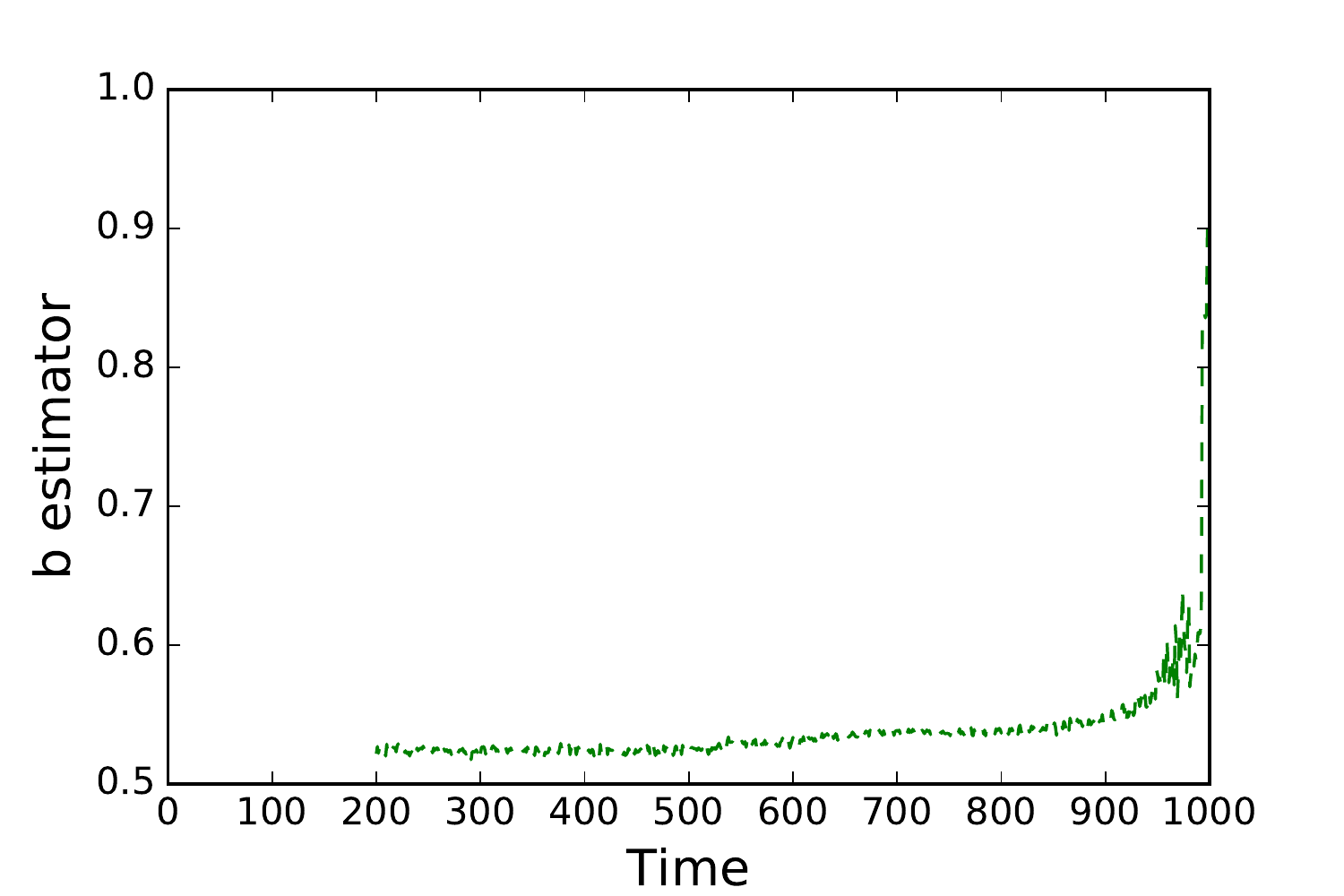}
}
\parbox{5cm}{\small \hspace{1.2cm}(a) ${\hat b}-t$ curve}
\end{minipage}
\hspace{0.2cm}
\begin{minipage}{4.1cm}
\centerline{
\includegraphics[width=1.9in]{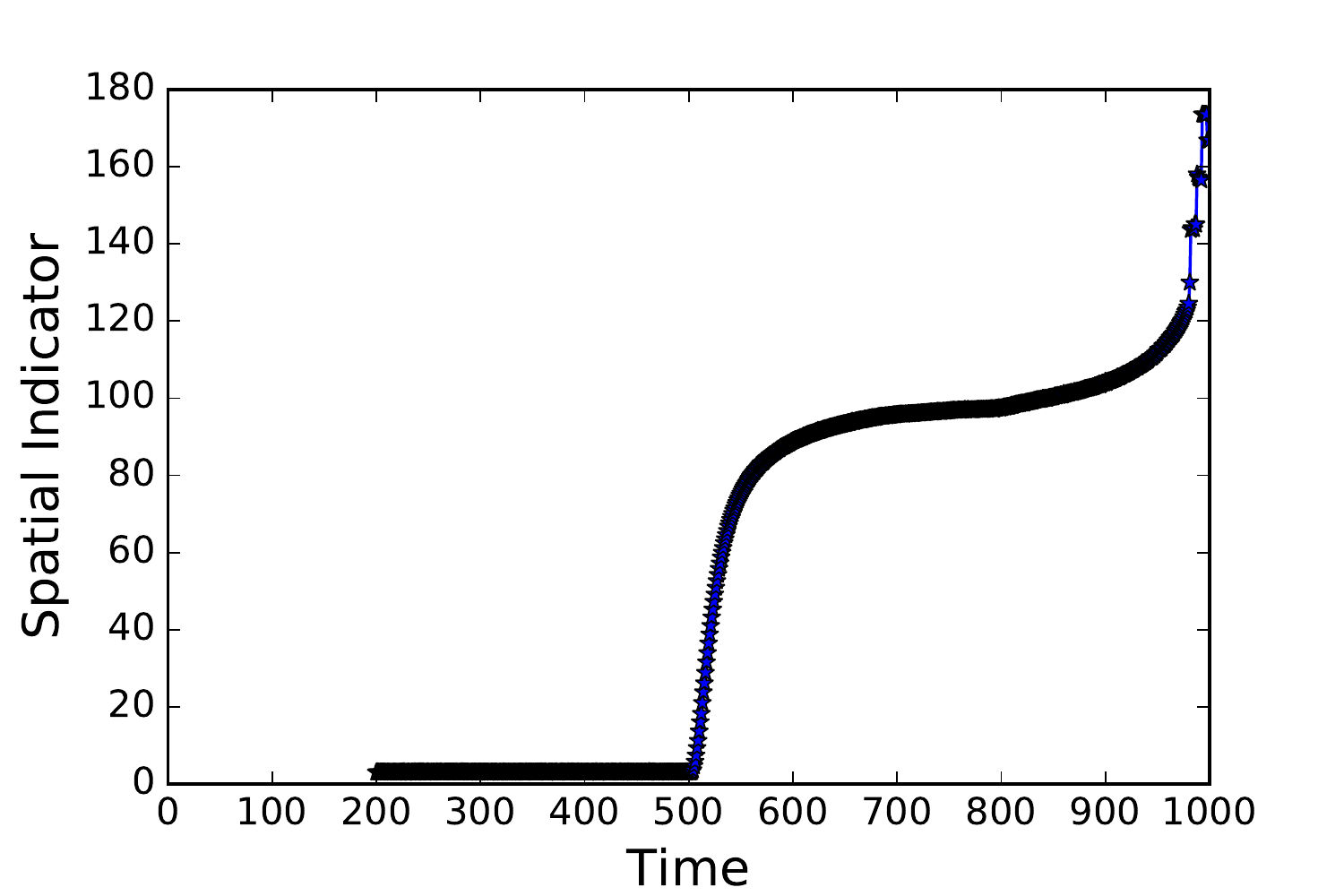}
}
\parbox{5cm}{\small \hspace{1.2cm}(b) $\mathcal{N}_{\phi}-t$ curve}
\end{minipage}
\hspace{0.2cm}
\caption{The increasing anomaly signal detection result in Case 2.}
\label{fig:b_case2_indicator}
\end{figure}
To illustrate the meaning of $b$, an increasing anomaly signal was set, which is shown in Table \ref{Tab: b_Case2}. The generated data is shown in Figure \ref{fig:b_case2_org}. The parameters were set the same as above. The generated ${\hat b}-t$ curve and $\mathcal{N}_{\phi}-t$ curve with continuously moving windows are shown in Figure \ref{fig:b_case2_indicator}. Interpretations of $b$ are stated as follows:

\uppercase\expandafter{\romannumeral1}. During $t_s=200\sim 500$, $\hat b$ and $\mathcal{N}_{\phi}$ remain almost constant, which indicates the system operates in normal state.

\uppercase\expandafter{\romannumeral2}. During $t_s=501\sim 979$, $\hat b$ increases gradually, which coincides with the variation of voltage caused by the gradually increasing signal. From $t_s = 980$, $\hat b$ begins to increase rapidly, which coincides with voltage collapse. It is noted that $\mathcal{N}_{\phi}$ increases rapidly since $t_s=501$, which indicates the spatial correlation of the residuals has been eliminated effectively. Combining the anomaly signal set in Table \ref{Tab: b_Case2}, it can be concluded that $\hat b$ is driven by the scale of anomaly signal.

3) Case Study on the Advantages of the Proposed Approach: In this case, by comparing with one-class support vector machines (SVMs) \cite{ma2003time}, structured autoencoders (AEs) \cite{liu2018anomaly}, long short term memory (LSTM) networks \cite{malhotra2015long}, and spectrum analysis (SA) based on the M-P law \cite{he2017big}, we validated the advantages of the proposed approach for anomaly detection, i.e., more sensitive to the variation of the spatio-temporal correlation in the data and robust to random fluctuations and measuring errors. The synthetic data generated in Figure \ref{fig:b_case2_org} was used to test the detection performances of different approaches. In the experiment, $SNR$ was set to be $200$. For SVMs, AEs and LSTM, we train the detection models only using a normal data sequence during $t_s=1\sim 200$ and compute the testing errors for the remaining sequence (i.e., $t_s=201\sim 1000$), in which one sampling data is used as a training/testing sample. The parameters involved in the proposed spatio-temporal analysis (STA) approach and the other methods are set as in Table \ref{Tab: Case3_parameter}.
\begin{table}[!t]
\caption{Parameter Settings Involved in the Detection Approaches.}
\label{Tab: Case3_parameter}
\centering
\footnotesize
\begin{tabular}{p{1.8cm}p{5.8cm}}   
\toprule[1.0pt]
\textbf{Approaches} & \textbf{Parameter Settings} \\
\hline
\multirow{2}*{SVMs} & the upper bound on the fraction of training errors $v$: 0.03; \\
~&the kernel function: $K({{\bf x}_i},{{\bf x}_j})={(0.01{{\bf x}_i}^{T}{{\bf x}_j})}^{3}$; \\
\hline
\multirow{7}*{AEs} & the model depth: $3$; \\
~&the number of neurons in each layer of encoder: $57,32,16$; \\
~&the number of neurons in each layer of decoder: $16,32,57$; \\
~&the initial learning rate: $0.001$;  \\
~&the activation function: $sigmoid$; \\
~&the minimum reconstruction error: $0.00001$;  \\
~&the optimizer: $Adam$.  \\
\hline
\multirow{7}*{LSTM} & the time steps: $1$; \\
~&the model depth: $3$; \\
~&the number of neurons in each layer: $57,64,57$; \\
~&the initial learning rate: $0.001$;  \\
~&the activation function: $sigmoid, tanh$; \\
~&the minimum reconstruction error: $0.00001$;  \\
~&the optimizer: $Adam$.  \\
\hline
\multirow{2}*{SA} & the moving window's size: $57\times 200$; \\
~&the test function: $\phi (\lambda)={\lambda}-log{\lambda}-1$. \\
\hline
\multirow{4}*{STA} & the moving window's size: $57\times 200$; \\
~&the test function: $\phi (\lambda)={\lambda}-log{\lambda}-1$; \\
~&the searching range of $p$: $1\sim 5$; \\
~&the searching step of $b$: $0.01$. \\
\hline
\end{tabular}
\end{table}

\begin{figure}[!t]
\centerline{
\includegraphics[width=3.0in]{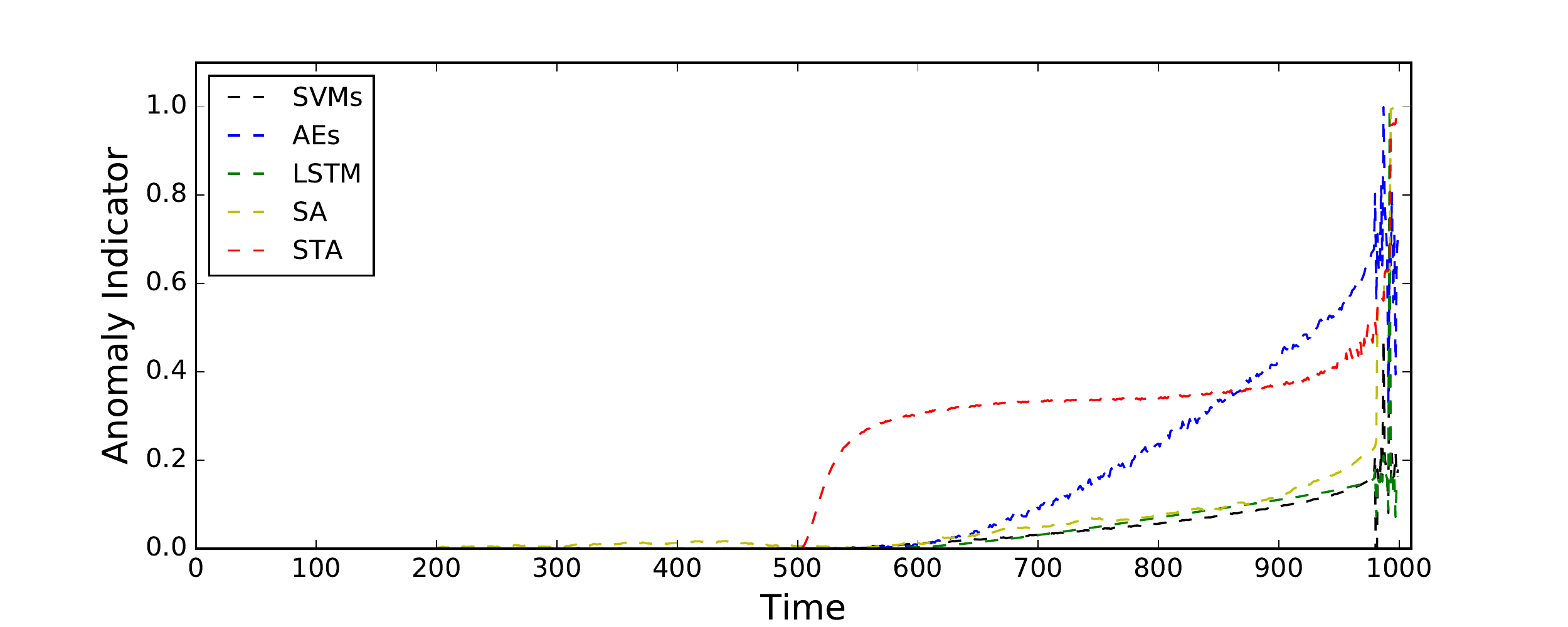}
}
\caption{Anomaly detection results of different approaches in Case 3.}
\label{fig:case3_comparison}
\end{figure}
The anomaly detection results of different approaches are normalized into $[0,1]$, which are shown in Figure \ref{fig:case3_comparison}. For SVMs, the normalization result of signed distance to the separating hyperplane is plotted; for AEs and LSTM, the normalized values of testing errors are plotted; for SA, the normalized value of linear eigenvalue statistics (LES) is plotted; for STA, the normalized value of $\mathcal{N}_{\phi}\times{\hat b}$ is plotted. Compared with the other approaches, STA is capable of detecting the anomaly signal much earlier (i.e., $t_s=501$) and easier, which indicates it is more sensitive to the anomaly signal and robust to the random fluctuations and measuring errors. The reason lies that, for each sampling time, a spatio-temporal data window instead of only the current sampling data is analyzed in the proposed approach. The average result makes the approach more robust to the random fluctuations and measuring errors.
\subsection{Case Study with Real-World Online Monitoring Data}
\label{section: case_real}
In this subsection, the online monitoring data obtained from a distribution network in Hangzhou city of China is used to validate the proposed approach. The distribution network contains $200$ feeder lines with $8000$ distribution transformers. For each feeder line, multiple online monitoring devices are installed and the online monitoring data are sampled every $15$ minutes. Anomaly information for each feeder line was recorded during the operation. In the following cases, three-phase voltages were chosen as the measurement variables to formulate the data matrices. Voltage disturbance was considered as the anomaly item.

\begin{figure}[!t]
\centering
\begin{minipage}{4.1cm}
\centerline{
\includegraphics[width=1.9in]{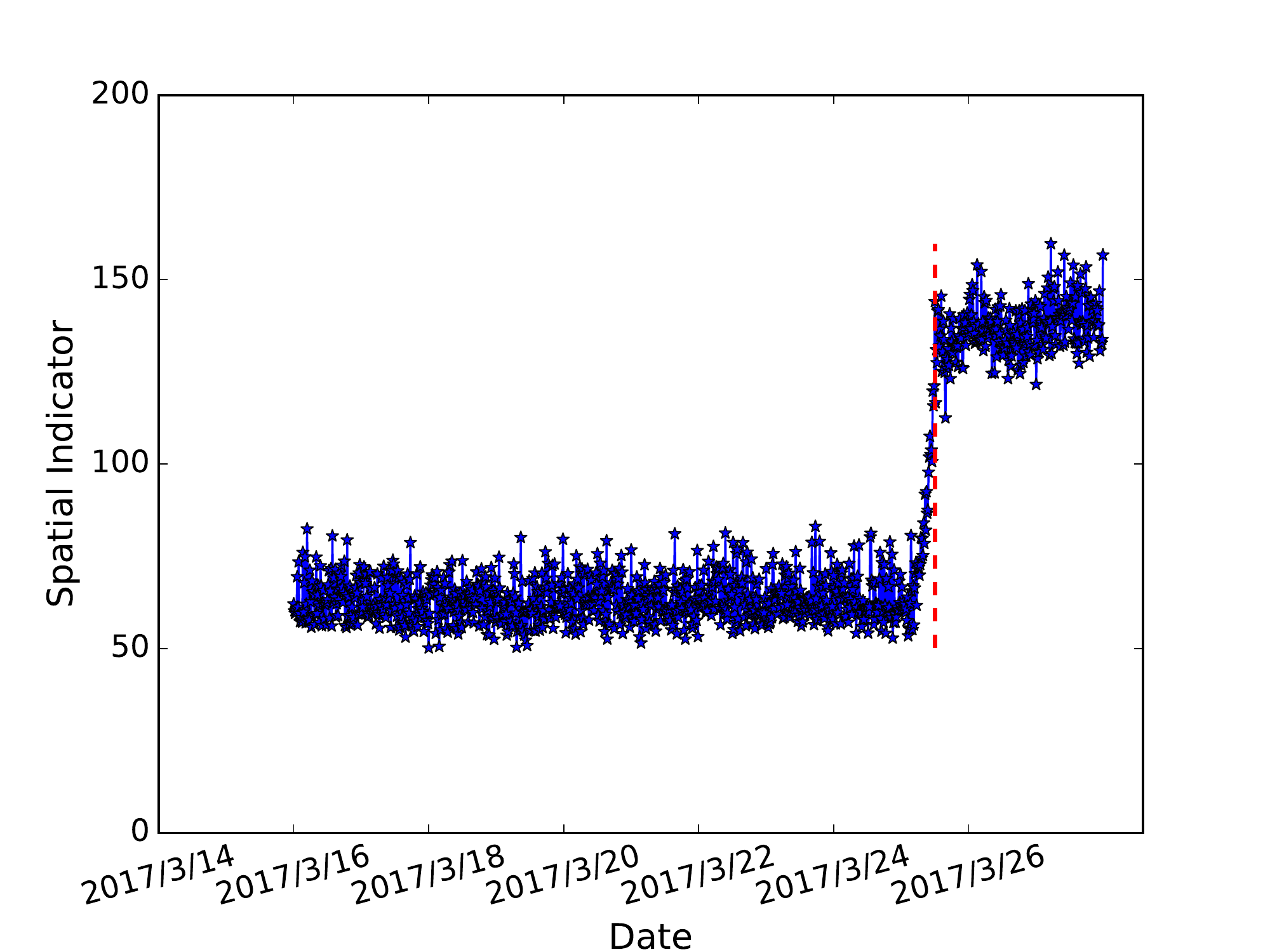}
}
\parbox{5cm}{\small \hspace{1.2cm}(a) $\mathcal{N}_{\phi}-t$ curve}
\end{minipage}
\hspace{0.2cm}
\begin{minipage}{4.1cm}
\centerline{
\includegraphics[width=1.9in]{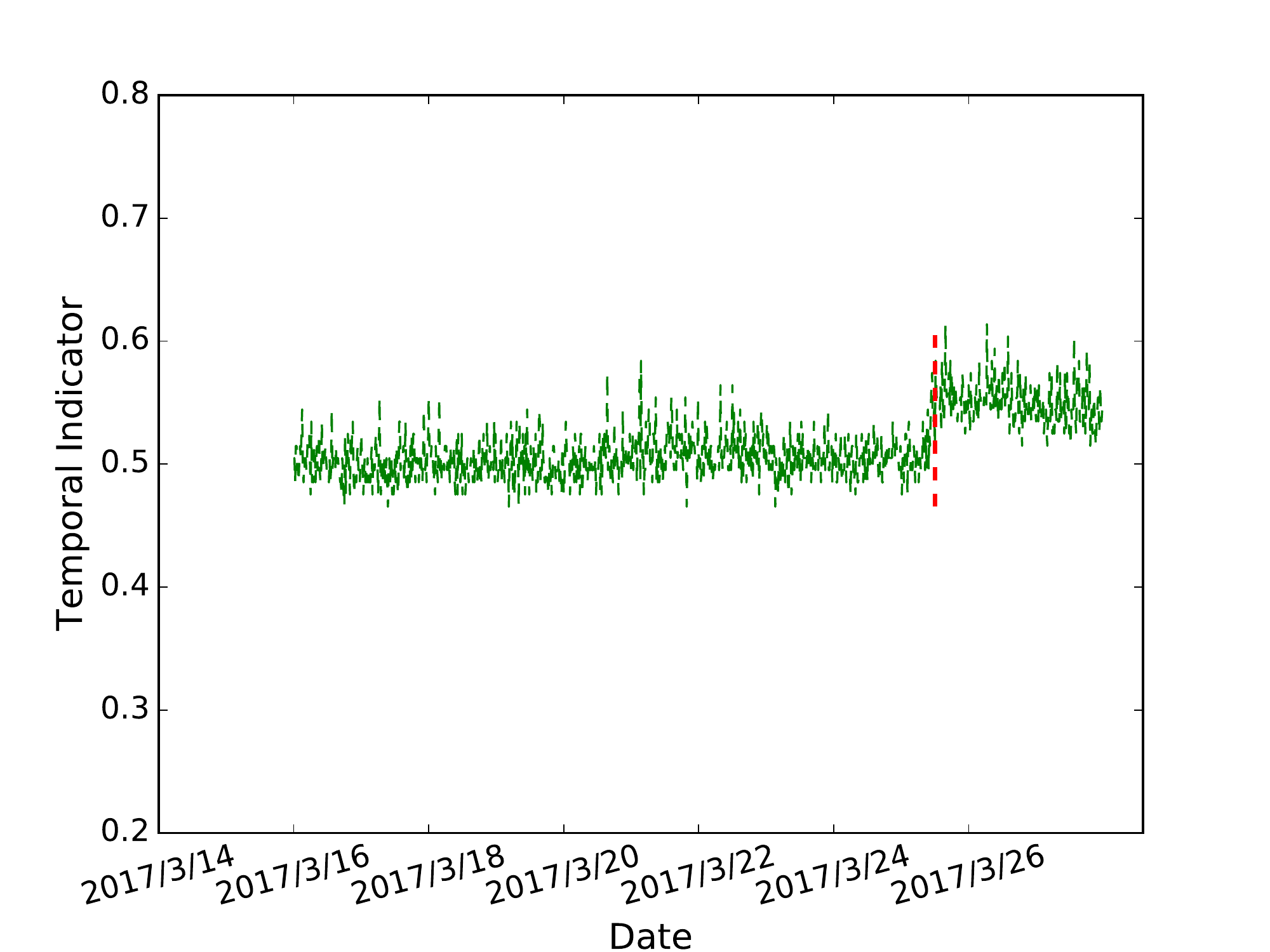}
}
\parbox{5cm}{\small \hspace{1.2cm}(b) ${\hat b}-t$ curve}
\end{minipage}
\caption{Effectiveness of our approach for voltage disturbance detection.}
\label{fig:casereal_indicator}
\end{figure}

4) Case Study on Voltage Disturbance: Voltage disturbance is an complex anomaly type in distribution networks, which may be caused by short circuit fault, sudden load change, or connection of distribution generation (DG), etc. In this case, we validated the effectiveness of the proposed approach by analyzing one feeder line suffering from voltage disturbance. $43$ online monitoring devices were installed on the feeder line and the researched data were sampled from 2017/3/14 00:00:00 to 2017/3/27 23:45:00, thus a $129\times 1344$ data matrix was formulated. The data with anomaly time and location information recorded are shown in Figure \ref{fig:casereal_org}. In the experiment, the moving window's size was set to be $129\times 192$. The generated $\mathcal{N}_{\phi}-t$ and $\hat b-t$ curves with continuously moving windows are shown in Figure \ref{fig:casereal_indicator}. In the figure, the red dashed line marks the beginning time of the anomaly. In calculating the confidence level $1-\alpha$ for each data point in the detection curves, $\mathcal{N}_{\phi}$ and $\hat b$ during continuous $672$ points ($671$ historical points and the current point) were considered to follow the student's t-distribution. The detection processes can be obtained as:

\begin{figure}[!t]
\centerline{
\includegraphics[width=3.0in]{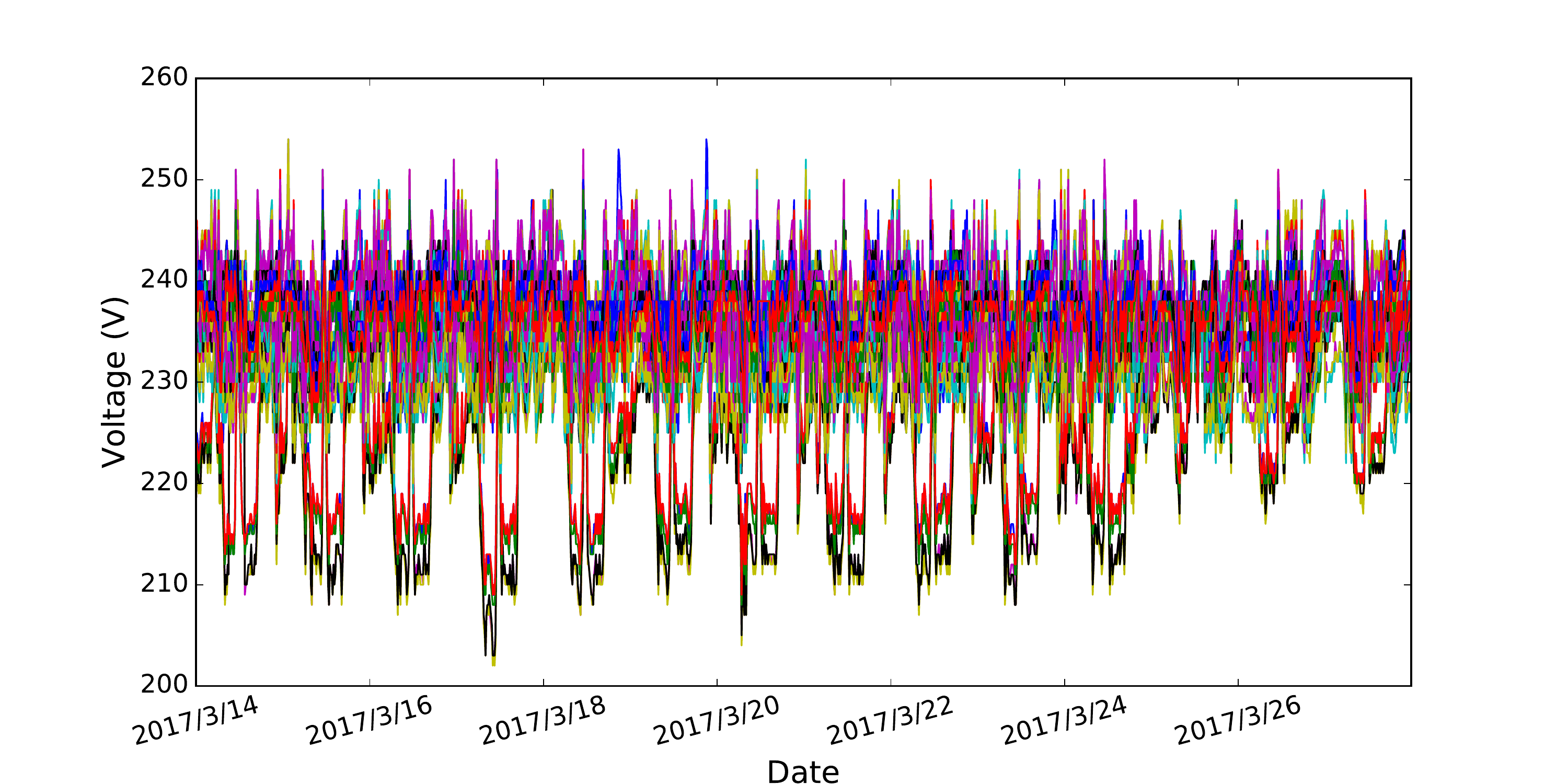}
}
\caption{The online monitoring data with anomaly time and index recorded. The anomaly time is 2017/3/25 12:00:00 and the anomaly index is $82\sim 87$.}
\label{fig:casereal_org}
\end{figure}
\begin{figure}[!t]
\centering
\begin{minipage}{4.1cm}
\centerline{
\includegraphics[width=1.9in]{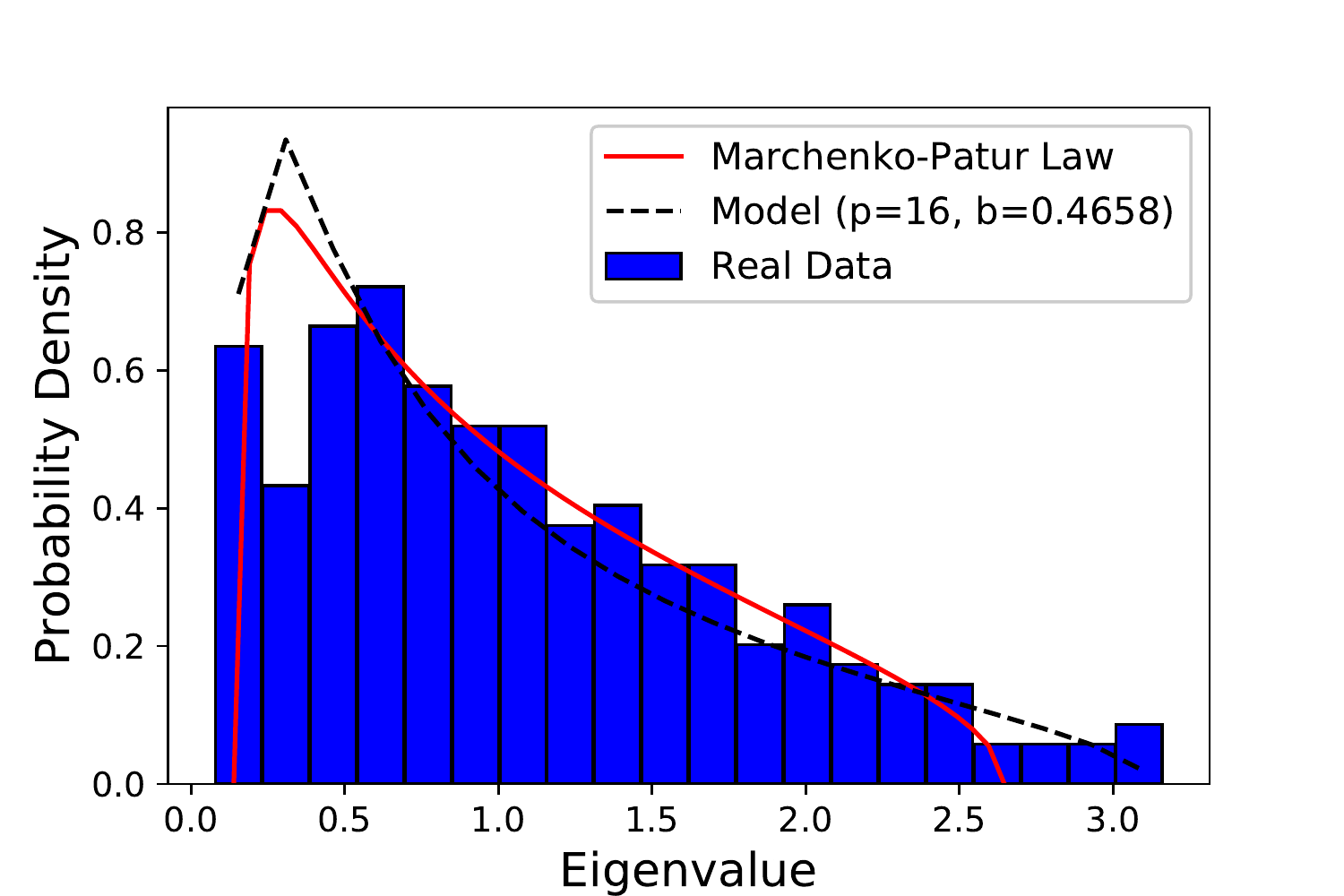}
}
\parbox{5cm}{\small \hspace{1.2cm}(a) Normal state}
\end{minipage}
\hspace{0.2cm}
\begin{minipage}{4.1cm}
\centerline{
\includegraphics[width=1.9in]{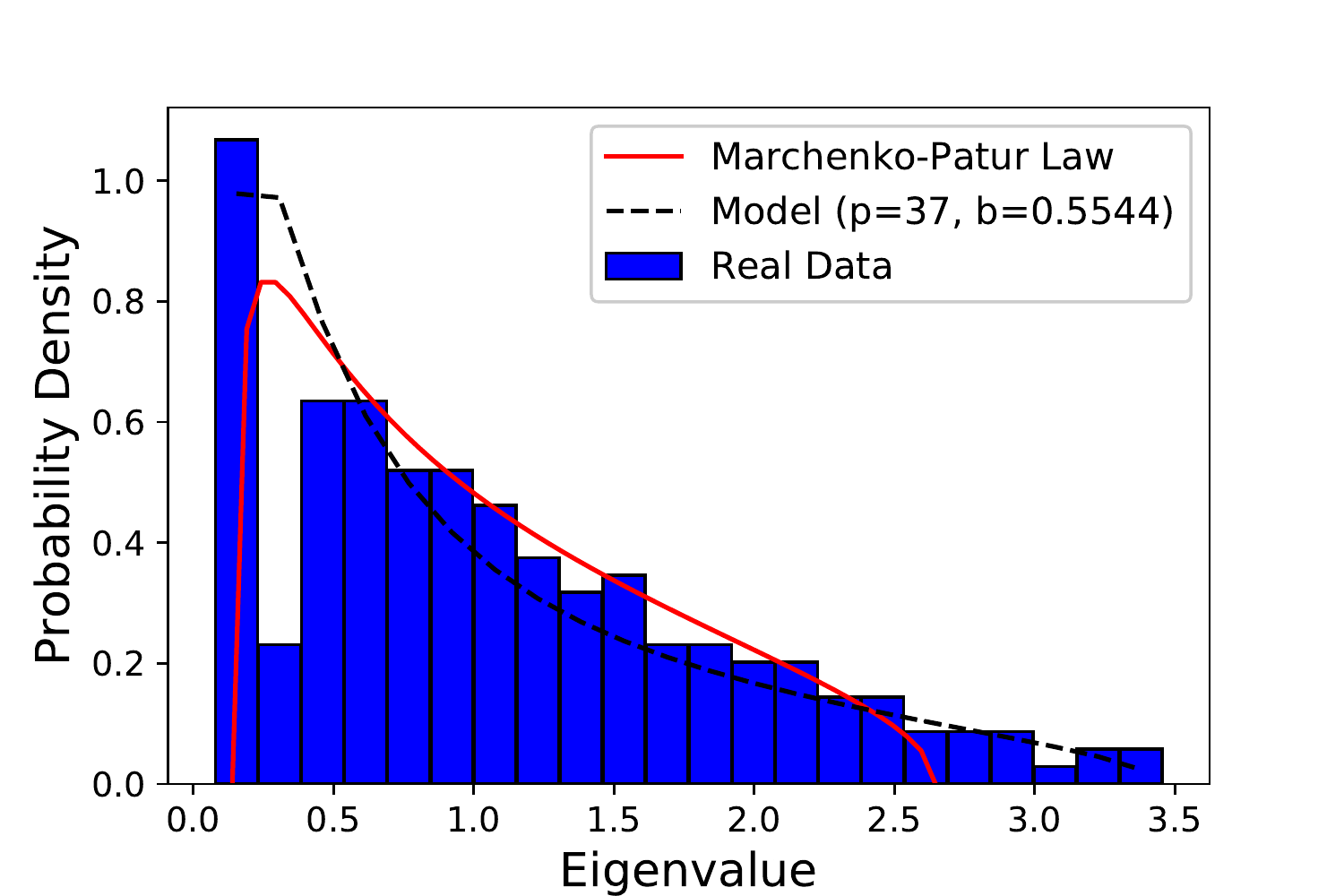}
}
\parbox{5cm}{\small \hspace{1.2cm}(b) Abnormal state}
\end{minipage}
\caption{The ESD of the covariance matrix of the residuals from the real-world online monitoring data can be fitted very well by our built model with the estimated $\hat p$ and $\hat b$, while can not be fitted by the M-P law. The optimal parameters estimated are different when the feeder operates in different states.}
\label{fig:casereal_esd}
\end{figure}

\uppercase\expandafter{\romannumeral1}. During 2017/3/14 00:00:00$\sim$2017/3/25 04:30:00, $\mathcal{N}_{\phi}$ and $\hat b$ remain almost constant and the values of $1-\alpha$ are small, which indicates the feeder line operates in normal state. For example, at 2017/3/25 04:30:00, the calculated $1-\alpha$ of $\mathcal{N}_{\phi}\times\hat b$ is $61.450\%$. As is shown in Figure \ref{fig:casereal_esd}(a), the ESD of covariance matrix of the residuals can be fitted well by the built model with ${\hat p=18, \hat b=0.4658}$ when the feeder line operates in normal state, but it does not fit the M-P law.

\uppercase\expandafter{\romannumeral2}. From 2017/3/25 04:45:00, $\mathcal{N}_{\phi}$ and $\hat b$ begin to change and the corresponding values of $1-\alpha$ increase rapidly, which indicates anomaly occurs and the operating state of the feeder line is becoming worse. For example, at 2017/3/25 04:45:00, the calculated $1-\alpha$ of $\mathcal{N}_{\phi}\times\hat b$ is $91.383\%$. Considering the recorded anomaly time is 2017/3/25 12:00:00, the proposed approach is able to detect the anomaly in an early phase. Figure \ref{fig:casereal_esd}(b) shows, in abnormal state, the ESD of covariance matrix of residuals can be fitted well by our built model with ${\hat p=37, \hat b=0.5544}$.

\begin{figure}[!t]
\centerline{
\includegraphics[width=3.0in]{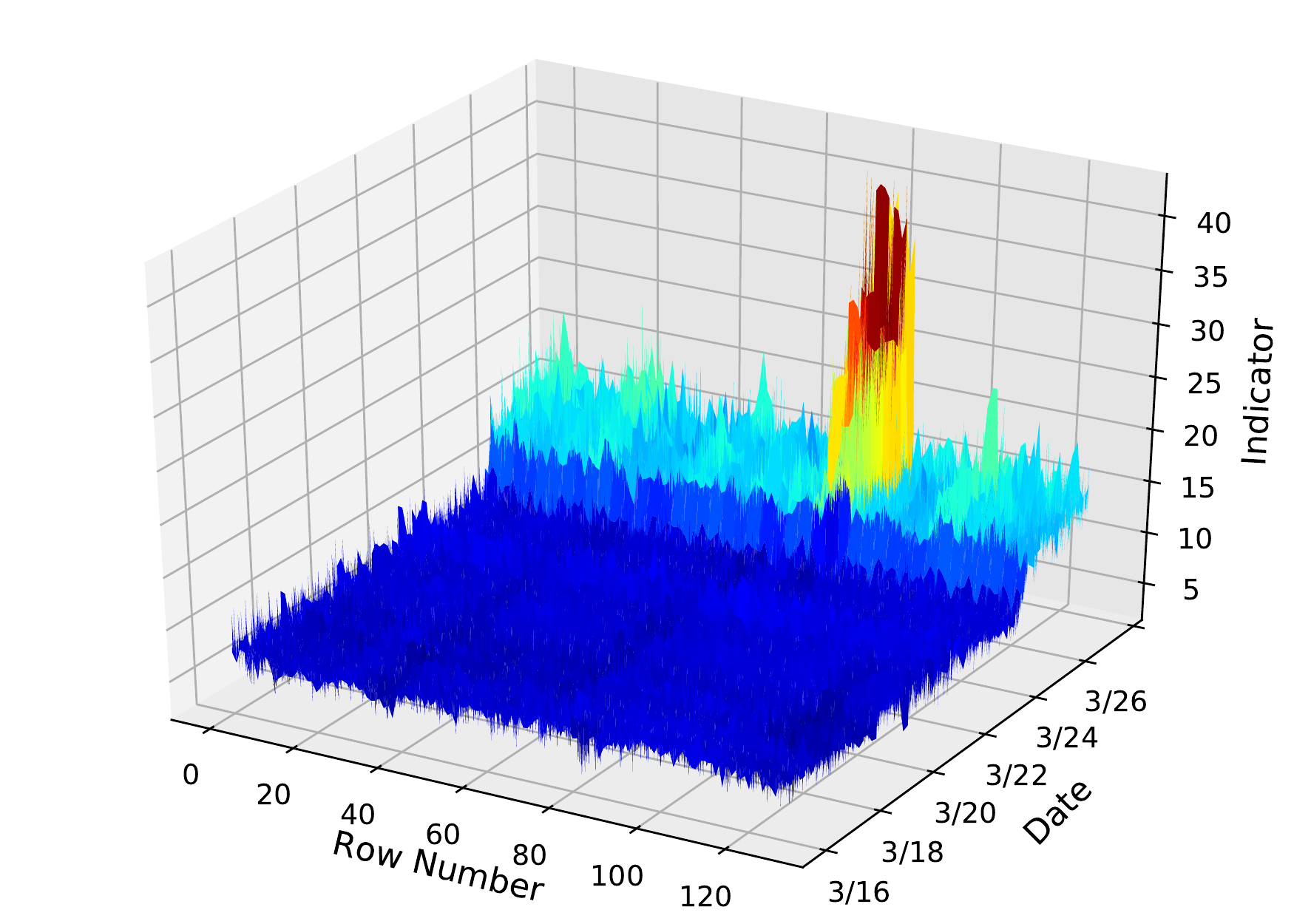}
}
\caption{Effectiveness of our approach for voltage disturbance location.}
\label{fig:casereal_loc}
\end{figure}
Furthermore, the anomaly is located, result of which is shown in Figure \ref{fig:casereal_loc}. It can be observed that, from 2017/3/25 04:45:00, the location indicator $\bm\eta$ increases rapidly and $\eta_{82\sim 87}$ are higher than others (such as $\eta_{80}$), which indicates the anomaly indexes are $82\sim 87$. For example, at 2017/3/25 04:45:00, the values of $1-\alpha$ for $\eta_{82\sim 87}$ and $\eta_{80}$ are $99.866\%, 99.682\%, 99.487\%, 99.927\%, 99.963\%, 99.975\%$, and $33.375\%$, respectively. The anomaly location results coincide with the recorded indexes in Figure \ref{fig:casereal_org}.

5) Case Study on Comparison with Other Approaches: In this case, we compare the proposed approach with one-class SVMs, structured  AEs, LSTM and SA based on the M-P law by detecting the anomalies in a distribution network. $180$ feeder lines with $80$ anomaly records during 2017/3/1 00:00:00$\sim$2017/3/28 23:45:00 were analyzed. The parameters involved in the detection approaches were set as in Table \ref{Tab: Case5_parameter}. For SVMs, AEs and LSTM, we trained the detection models using 7 days' normal data sequence and computed the testing errors for the sequence to be analyzed; for SA, the LES was calculated for each moving window; for STA, $\mathcal{N}_{\phi}\times{\hat b}$ was calculated for each moving window. The value of  $1-\alpha$ for each data point in the detection curves was calculated and ${(1-\alpha)}_{th}$ was set as $95\%$.
\begin{table}[!t]
\caption{Parameter Settings Involved in the Detection Approaches.}
\label{Tab: Case5_parameter}
\centering
\footnotesize
\begin{tabular}{p{1.8cm}p{5.8cm}}   
\toprule[1.0pt]
\textbf{Approaches} & \textbf{Parameter Setting} \\
\hline
\multirow{3}*{SVMs} & the upper bound on the fraction of training errors $v$: 0.1; \\
~&the kernel function: \\
~&$K({{\bf x}_i},{{\bf x}_j})=exp{({-\frac{1}{2N}}{||{{\bf x}_i}-{{\bf x}_j}||}^{2})}$; \\
\hline
\multirow{7}*{AEs} & the model depth: $3$; \\
~&the number of neurons in each layer of encoder: $\lfloor\{1,0.6,0.3\}\times N\rfloor$; \\
~&the number of neurons in each layer of decoder: $\lfloor\{0.3,0.6,1\}\times N\rfloor$; \\
~&the initial learning rate: $0.001$;  \\
~&the activation function: $sigmoid$; \\
~&the minimum reconstruction error: $0.00001$;  \\
~&the optimizer: $Adam$.  \\
\hline
\multirow{7}*{LSTM} & the time steps: $96$; \\
~&the model depth: $3$; \\
~&the number of neurons in each layer: $\{1,1,1\}\times N$; \\
~&the initial learning rate: $0.001$;  \\
~&the activation function: $sigmoid, tanh$; \\
~&the minimum reconstruction error: $0.00001$;  \\
~&the optimizer: $Adam$.  \\
\hline
\multirow{2}*{SA} & the moving window's size: $N\times 192$; \\
~&the test function: $\phi (\lambda)={\lambda}-log{\lambda}-1$. \\
\hline
\multirow{4}*{STA} & the moving window's size: $N\times 192$; \\
~&the test function: $\phi (\lambda)={\lambda}-log{\lambda}-1$; \\
~&the searching range of $p$: $1\sim\lfloor\frac{N}{2}\rfloor$; \\
~&the searching step of $b$: $0.01$. \\
\hline
\end{tabular}
\end{table}

To compare the detection performances of different approaches, we use the $true\; detecting\; rate\; (TDR)$ and $false\; alarming\; rate\; (FAR)$ to measure the performance of each approach. The $TDR$ and $FAR$ are defined as
\begin{equation}
\label{Eq:performance_detection}
\begin{aligned}
  &TDR = \frac{N_{cr}}{N_{gt}} \\
  &FAR = \frac{N_{al}-N_{cr}}{N_{al}}
\end{aligned},
\end{equation}
where $N_{cr}$ is the number of anomalies that are correctly detected, $N_{gt}$ denotes the number of ground-truth anomalies, and $N_{al}$ is the number of all detected alarms. The higher the $TDR$ and the smaller the $FAR$, the better detection performance of an approach. Meanwhile, in order to compare the efficiency of different approaches, the $average\; calculation\; time\;(ACT)$ for each sampling time was counted. For SVMs, AEs and LSTM, the ACT for each testing sample was counted, which does not include the model training time. The experiments were conducted on a server with $2.60$ GHz central processing unit (CPU) and $8.00$ GB random access memory (RAM). The comparison results are shown in Table \ref{Tab: Case5_comparison}.
\begin{table}[!t]
\caption{Comparison Results of Different Detection Approaches.}
\label{Tab: Case5_comparison}
\centering
\footnotesize
\begin{tabular}{cccc}   
\toprule[1.0pt]
\textbf{Approaches} & $\bm{TDR}$($\%$) & $\bm{FAR}$($\%$) & $\bm{ACT}$(s) \\
\hline
SVMs & 65.00 & 45.83 & 0.0012  \\
AEs & 86.25 & 21.59 & 0.024 \\
LSTM & 77.50 & 27.91 & 0.087  \\
SA & 70.00 & 30.86 & 0.790 \\
STA & 85.00 & 16.04 & 3.326 \\
\hline
\end{tabular}
\end{table}

From Table \ref{Tab: Case5_comparison}, it can be observed that structured AEs and STA outperform the other approaches in anomaly detection accuracy. It is noted that STA has the smallest $FAR$, which indicates it is more robust to random fluctuations and measuring errors in the data. Meanwhile, it can be seen that our proposed approach has the highest $ACT$ for the reason of searching $p$ and $b$ with minimal step size. In practice, the efficiency of the proposed approach can be improved by restricting the searching ranges empirically and using a larger searching step size. Considering that the online monitoring data in the researched network are sampled every $15$ minutes, the proposed approach is practical for online data analysis. Compared with SVMs, structured AEs and LSTM, STA is an unsupervised approach and it does not rely on any labels. Compared with SA based on the M-P law, STA is more accurate in dissecting the complex spectrum of the real data, which makes it more sensitive to the variation of the correlation in the data.
\section{Conclusion}
\label{section: conclusion}
By analyzing the structure information of the online monitoring data in distribution networks, a spatio-temporal correlation analysis approach is proposed for anomaly detection and location in this paper. It is capable of detecting the anomalies in an early phase by exploring the variation of the spatio-temporal correlation in the data. The spatial and temporal indicators we designed are able to indicate the data behaviour accurately. The proposed approach is purely data-driven and it does not require prior knowledge on the complex topology of the distribution network. It is robust to random fluctuations or measuring errors in the data, which can help reduce the false alarming rate. The case studies with synthetic data verify the effectiveness and advantages of the proposed approach and offer explanations on the involved spatio-temporal parameters. Through the real-world online monitoring data from a distribution network, we validate the approach and compare it with the other existing techniques. The results show the advantages of the proposed approach for anomaly detection and location, and it can be served as a primitive for analyzing the spatio-temporal data in distribution networks.

\appendices
\section{Marchenko-Pastur Law}
\label{section: mp_law}
Let ${\bf X}=\{{x}_{i,j}\}$ be a $N \times T$ random matrix, whose entries are independent identically distributed (i.i.d.) variables with the mean $\mu (x)=0$ and the variance $\sigma ^2 (x)<\infty$. The corresponding covariance matrix is defined as ${\bf \Sigma}=\frac{1}{T} {\bf X}{\bf X}^{H}$. As $N,T \to\infty$ but $c=\frac{N}{T}\in (0,1]$, according to the M-P law, the ESD of ${\bf\Sigma}$ converges to the limit with probability density function (PDF)
\begin{equation}
\label{Eq:mp-law}
\begin{aligned}
{f_{MP}}(x) = \left\{ \begin{array}{l}
\frac{1}{{2\pi c{\sigma ^2}}x}\sqrt {(b - x)(x - a)} {\rm{,}} \quad a \le x \le b\\
0, \qquad  \qquad  \qquad  \qquad  \qquad  \quad {\rm{others}}
\end{array} \right.
\end{aligned},
\end{equation}
where $a={\sigma ^2}{(1-\sqrt{c})}^2$, $b={\sigma ^2}{(1+\sqrt{c})}^2$.

\section{Derivation Details of the Polynomial Equation}
\label{section: derivation}
\begin{definition}
\emph{The Green's Function} (\emph{or Stieltjes Transform}).
\begin{equation}
\label{Eq:stieltjes transform}
\begin{aligned}
  G_H(z)=\int\frac{\rho_H(\lambda)}{z-\lambda}d\lambda
\end{aligned},
\end{equation}
where $\rho_H(\lambda)$ is the spectral density (i.e., the eigenvalue density) of the random matrix $\bf H$, which can be reconstructed from the Green's Function by calculating its imaginary part
\begin{equation}
\label{Eq:spectral_density}
\begin{aligned}
  \rho_H(\lambda)=-\frac{1}{\pi} \lim\limits_{\varepsilon\rightarrow 0^{+}}\Im G_{\Sigma_{model}}(\lambda + i\varepsilon)
\end{aligned}.
\end{equation}
\end{definition}

\begin{definition}
\emph{Moment}.

The $n$-th moment of $\rho_H(\lambda)$ is defined as
\begin{equation}
\label{Eq:moment}
\begin{aligned}
  m_{H,n}=\int\rho_H(\lambda){\lambda}^nd\lambda
\end{aligned}.
\end{equation}
\end{definition}

\begin{definition}
\emph{Moment generating function}.

\begin{equation}
\label{Eq:moment_generating_function_G}
\begin{aligned}
  G_H(z)=\sum\limits_{n=0}^{\infty}\frac{m_{H,n}}{z^{n+1}}
\end{aligned}.
\end{equation}
\begin{equation}
\label{Eq:moment_generating_function_M}
\begin{aligned}
  M_H(z)=\sum\limits_{n=1}^{\infty}\frac{m_{H,n}}{z^{n+1}}
\end{aligned}.
\end{equation}
Thus, the relation between $M_H(z)$ and $G_H(z)$ can be derived through equation (\ref{Eq:moment_generating_function_G}) and (\ref{Eq:moment_generating_function_M})
\begin{equation}
\label{Eq:relation_M_G}
\begin{aligned}
  M_H(z)=zG_H(z)-1
\end{aligned}.
\end{equation}
\end{definition}

\begin{definition}
\emph{N-transform}.

$N_H(z)$ is the inverse transform of $M_H(z)$, namely,
\begin{equation}
\label{Eq:relation_M_N}
\begin{aligned}
  M_H(N_H(z))=N_H(M_H(z))=z
\end{aligned}.
\end{equation}
\end{definition}

For the empirical covariance matrix $\Sigma = \frac{1}{T}UU^{T} = \frac{1}{T}{A_{N}^{1/2}}S{B_T}{S^T}{A_{N}^{1/2}}$, the N-transform of $\Sigma$ can be derived as
\begin{equation}
\label{Eq:N_transform}
\begin{aligned}
  N_\Sigma(z)&=N_{\frac{1}{T}{A_{N}^{1/2}}S{B_T}{S^T}{A_{N}^{1/2}}}(z) \\
  & =N_{\frac{1}{T}S{B_T}{S^T}{A_N}}(z) \; ( cyclic\; property\;of\; trace) \\
  & =\frac{z}{1+z}N_{\frac{1}{T}S{B_T}{S^T}}(z)N_{A_N}(z) \;(FRV\;multiplication\;law) \\
  & =\frac{z}{1+z}N_{\frac{1}{T}{S^T}S{B_T}}(rz)N_{B_T}(z) \;(cyclic\; property\;of\; trace) \\
  & =\frac{z}{1+z}\frac{rz}{1+rz}N_{\frac{1}{T}{S^T}S}(rz)N_{B_T}(rz)N_{A_N}(z) \;(FRV) \\
  & =rzN_{B_T}(rz)N_{A_N}(z)
\end{aligned}.
\end{equation}
Considering $M\equiv M_\Sigma(z)$ and its inverse relation to N-transform, we can obtain
\begin{equation}
\label{Eq:key_equation}
\begin{aligned}
  z=rMN_{B_T}(rM)N_{A_N}(M)
\end{aligned}.
\end{equation}

In Section \ref{section: residual}, we assume the cross-correlations of the real residuals are effectively eliminated by removing factors. Thus, the cross-correlation matrix $A_N=I_N$, and $N_{A_N}(z)=N_{I_N}(z)=1+1/z$. By combining equation (\ref{Eq:key_equation}), we can obtain
\begin{equation}
\label{Eq:key_equation_AR}
\begin{aligned}
  z=r(1+M)N_{B_T}(rM) \\
  \Updownarrow\qquad\qquad \\
  rM=M_{B_T}(\frac{z}{r(1+M)})
\end{aligned}.
\end{equation}

In Section \ref{section: residual}, we assume the auto-correlations of the real residuals follow an autoregressive process, thus $\{{B_T}\}_{it}=|b|^{i-t}$. By using Fourier-transform, the moment generating function of $B_T$ is given by
\begin{equation}
\label{Eq:moment_generating_function_B}
\begin{aligned}
  M_{B_T}(z)=-\frac{1}{\sqrt{1-z}\sqrt{1-\frac{(1+b^2)^2}{1-b^2}z}}
\end{aligned}.
\end{equation}
Thus, by combining equation (\ref{Eq:key_equation_AR}) and (\ref{Eq:moment_generating_function_B}), we can obtain the object polynomial in equation (\ref{Eq:polynomial}).

\section{Proof for Anomaly Location}
\label{section: proof}

Let $\hat F$ and $\bm\lambda$ be the principal components and corresponding eigenvalues from the covariance matrix $C=\frac{1}{T}{R}{R}^{H}$, where $R$ is an $N\times T$ real matrix. According to the matrix theory, we can obtain
\begin{equation}
\label{Eq:covariance_C}
\begin{aligned}
  C{\hat F}^{(i)} = \lambda_{i}{\hat F}^{(i)}
\end{aligned}.
\end{equation}
The derivation of equation (\ref{Eq:covariance_C}) regarding its entries $c_{jk} (j,k=1,\cdots,N)$ is
\begin{equation}
\label{Eq:derivation_C}
\begin{aligned}
  \frac{dC}{dc_{jk}}{\hat F}^{(i)}+C\frac{d{\hat F}^{(i)}}{dc_{jk}} = \frac{d\lambda_{i}}{dc_{jk}}{\hat F}^{(i)}+\lambda_{i}\frac{d{\hat F}^{(i)}}{dc_{jk}}
\end{aligned}.
\end{equation}
Since $C$ is real and symmetric, there exists ${({\hat F}^{(i)})}^{H}{\hat F}^{(i)} = 1$. Left multiply ${({\hat F}^{(i)})}^{H}$ for equation (\ref{Eq:derivation_C}), we can obtain
\begin{equation}
\label{Eq:derivation_C_sim1}
\begin{aligned}
  \frac{d\lambda_{i}}{dc_{jk}} = {({\hat F}^{(i)})}^{H}\frac{dC}{dc_{jk}}{\hat F}^{(i)}
\end{aligned},
\end{equation}
where $\frac{d\lambda_{i}}{dc_{jk}}$ gets the value of $1$ only for the entry $c_{jk}$ in $C$ and $0$ for the others. Thus, equation $(\ref{Eq:derivation_C_sim1})$ can be simplified as
\begin{equation}
\label{Eq:derivation_C_sim2}
\begin{aligned}
  \frac{d\lambda_{i}}{dc_{jk}} = {{\hat F}^{(i,j)}}{\hat F}^{(i,k)}
\end{aligned},
\end{equation}
where ${{\hat F}^{(i,j)}}$ and ${\hat F}^{(i,k)}$ represent the $j-$th and $k-$th element of the principal component ${\hat F}^{(i)}$. Then the contribution of the $j-$th row's elements to $\lambda_{i}$ can be measured by
\begin{equation}
\label{Eq:derivation_C_sim3}
\begin{aligned}
  \sum\limits_{k=1}^{N}{(\frac{d\lambda_{i}}{dc_{jk}})}^2 = {({\hat F}^{(i,j)})}^2\sum\limits_{k=1}^{N}{({\hat F}^{(i,k)})}^2 = {({\hat F}^{(i,j)})}^2
\end{aligned}.
\end{equation}

\small{}
\bibliographystyle{IEEEtran}
\bibliography{helx}

\normalsize{}
\end{document}